\documentclass[review,3p,sort&compress]{elsarticle}
\usepackage{amsmath}
\usepackage{amssymb}
\usepackage{amsthm}
\usepackage{mathtools}
\usepackage{graphicx}% Include figure files
\usepackage{xcolor}
\DeclareMathOperator{\sgn}{sgn}
\usepackage[font={small,stretch=0.95},figurename={Fig.},labelfont=bf,labelsep=period]{caption}
\usepackage[colorlinks=true]{hyperref}
\usepackage{multirow}
\usepackage{array}

\journal{ }

\begin{document}
	
	\begin{frontmatter}

		\title{Higher-order topological corner states and edge states in grid-like frames}
		
		\author[aff1]{Yimeng Sun\fnref{equal}}
		%\ead{sym@stu.pku.edu.cn}
		
		\author[aff1]{Jiacheng Xing\fnref{equal}}
		
		\author[aff2]{Li-Hua Shao}
		
		\author[aff1]{Jianxiang Wang\corref{cor}}
		\ead{jxwang@pku.edu.cn}
		
		\cortext[cor]{Corresponding author.}
		
		\fntext[equal]{These authors contributed equally to this work.}
		
		\affiliation[aff1]{organization={Department of Mechanics, School of Mechanics and Engineering Science, Peking University},
			city={Beijing},
			postcode={100871},
			country={China}}
		
		\affiliation[aff2]{organization={School of Aeronautic Science and Engineering, Beihang University},
			city={Beijing},
			postcode={100191},
			country={China}}
		
		\begin{abstract}
			Continuum grid-like frames composed of rigidly jointed beams are classic subjects in the field of structural mechanics, whose topological dynamical properties have only recently been revealed. For two-dimensional frames, higher-order topological phenomena may occur, with frequency ranges of topological states and bulk bands becoming overlapped, leading to hybrid mode shapes. Concise theoretical results are necessary to identify the topological modes in such planar continuum systems with complex spectra. In this work, we present analytical expressions for the frequencies of higher-order topological corner states, edge states, and bulk states in kagome frames and square frames, as well as the criteria of existence of these topological states and patterns of their distribution in the spectrum. We identify the edge and corner states even under their degeneracy with the bulk bands. We show that the corner states are within the bandgaps of edge states unless topological transitions occur, and demonstrate the robustness of higher-order topological states under perturbations. These theoretical results fully demonstrate that the grid-like frames, despite being a large class of two-dimensional continuum systems, have topological states that can be accurately characterized through concise analytical expressions. This work contributes to the study of topological mechanics, and the accurate and concise theoretical results facilitate direct applications of topological grid-like frame structures in industry and engineering.

		\end{abstract}
		
		
	\end{frontmatter}

	\section{Introduction}
	Topological mechanical materials \citep{huber_topological_2016}, originating from topological insulators in condensed-matter physics \citep{von_klitzing_40_2020}, can host topologically protected mechanical modes that exhibit robustness, and have received widespread attention in engineering fields. Although topological mechanical materials share many similarities with topological insulators in physical mechanisms, constructing novel topological mechanical systems requires appropriate formulations that are based on mechanical principles and consequently admit topological properties mathematically. In the field of topological mechanics, spring--mass systems are widely adopted to realize static topological zero modes \citep{kane_topological_2014,chen_observation_2025}, modes of self-stress \citep{kane_topological_2014}, and dynamical topological modes \citep{Springmass-Prl,Chan2018,chen2018,chen2019,Shisd2021,YANG2024105889}, usually enabling an analytical description of the topological properties \citep{chen2019,wang_static_2026}. On the other hand, continuum topological mechanical systems \citep{liu_topological_continuum} have received growing interest due to the wider applicability of continuum materials in real-world scenarios; in this respect, truss structures are tailored to host static topological modes \citep{annurev-mao,Floppymode}, and continuum elastic structures with spatially modulated stiffness or mass are constructed to achieve topological modes in the dynamical regime \citep{yin_band_2018,MUHAMMAD2019359,PRLpumping,fan2019,Zhou2020,zhang_multiple_2025}. Research in this regard has greatly facilitated the application of topological modes in exploring novel wave phenomena \citep{zhang_dirac_2020,CHENG2024105475}. Typically, the analysis of topological modes in continuum elastic systems is largely based on numerical calculations such as finite-element simulations or experimental methods \citep{wu_-plane_2020,DUAN2023105251,acta}. However, the infinite-degree-of-freedom nature of the continuum endows such systems with not only multiple frequency bands but also multiple topological phase transition points \citep{xiao_surface_2014,yin_band_2018,MUHAMMAD2019359}, for which it is challenging to obtain a relation between the topological phases of different bands and the structural parameters in a simple and exact form. The characterization of topological properties of continuum systems in an analytical manner and identification of topological phase transitions across the whole frequency spectrum are desired for fully revealing and applying the ample topological properties of continuum systems. To this end, the authors have recently revealed the topological dynamical properties for a class of one-dimensional (1D) continuous beam structures, and set up a theoretical framework to describe the multiple topological phase transitions therein \citep{SUN2025105935}. We demonstrated that the theoretical framework can be applied to reveal the topological
dynamics of a broad class of continuum grid-like frame structures such as square frames, kagome frames, and bridge-like frames, and some topological modes at certain frequencies are demonstrated as examples \citep{SUN2025105935}.
	
	Compared with conventional topological insulators, higher-order topological insulators can accommodate not only robust edge/surface states, but also lower-dimensional corner states \citep{doi:10.1126/science.aah6442,doi:10.1126/sciadv.aat0346,PhysRevLett.120.026801,PhysRevB.98.205147,Benalcazar2019,xie_higher-order_2021,PhysRevLett.132.197202,ling_twist-induced_2026}. Topological mechanical systems, due to the advantage of macroscopic size and flexibility in frequency excitations \citep{ma_topological_2019}, have also become a popular platform for research in higher-order topological phases \citep{fan2019,PRLpumping,wu_-chip_2021}. Higher-order topological materials feature more intricate band topology and phase transitions, and may commonly be accompanied by degeneracy between different types of modes \citep{liu_novel_2017,zhang_second-order_2019,obana_topological_2019}. Integrating higher-order topology into continuum mechanical systems would pose a notable challenge as to an exact characterization of the topological modes and phase transitions in an analytical manner. Grid-like frames \citep{HUYBRECHTS19961001,WANG1998417,10.1093/qjmam/56.1.45,HUNT2022115120} are structures composed of intersecting elastic beams; while such structures do not exhibit an apparent ``concentrated mass'' characteristic \citep{wu_-plane_2020,chen_corner_2021,wang_elastic_2021,yi_delocalization_2024}, they provide an effective and practical platform for accommodating higher-order topological phases, being both easy to manufacture and omnipresent in reality in various forms. In two-dimensional (2D) grid-like frames, as the frequency ranges of higher-order corner states and edge states may become overlapped with bulk states, leading to possible hybridization of the mode shapes, it is difficult to exactly identify the topological modes through numerical techniques. Therefore, an analytical approach able to derive the existence and frequency ranges of topological states for grid-like frames is of crucial importance for us to gain an understanding of the topological phases of higher-order topological continuum systems and guide practical engineering applications.
	
	In this paper, we present an analytical method to characterize the higher-order topological properties of planar continuum grid-like frames. For square and kagome frames, we derive exact analytical expressions for topological phase transitions in terms of geometric parameters. Although the topological corner states, edge states, and bulk states in higher-order topological grid-like frames may have overlapping frequency regions, the proposed analytical method enables determination of exact frequency ranges for the corner, edge, and bulk states, as well as clear existence conditions of the topological modes. Furthermore, through both theoretical arguments and finite-element simulations, we demonstrate that while higher-order topological corner states may be degenerate with bulk states, they remain localized within the bandgap of edge states and retain robustness. We extend the theory to topological heterostructures, illustrating that connecting grid-like frames with different higher-order topological phases results in the emergence of topological corner states at the interface. The concise theoretical results on the topological dynamics of complex frame structures facilitate direct applications of these findings to industrial engineering domains such as safety assessment and robust waveguiding \citep{phani_wave_2006,advs.201900401}.

	\section{Preliminaries}
	The authors have recently presented a theoretical framework for the band-structure analysis of a broad class of continuum elastic frame structures, along with a method of deriving the dynamical matrix \citep{SUN2025105935}. For present purposes, we are concerned with the in-plane motion of several types of frame structures consisting of beams with uniform linear density $m$ and bending stiffness $EI$, wherein the following rules apply:
	\begin{itemize}
		\item The governing equation for harmonic modes is $H \lvert\theta\rangle = 0$, where $H$ is a square matrix, and $\lvert\theta\rangle$ comprises the rotation angles at all joints (i.e., intersection points of beam segments) of the frame structure\footnote{The frame structures concerned in this paper (square and kagome frames) share the crucial property that every single joint cannot translate with the beams undergoing no elongation. This results in only rotational degrees-of-freedom at joints.}.
		\item The ``dynamical matrix'' $H$ relies on the frequency $\omega$, with matrix elements being analytical functions of $\beta = \sqrt[4]{\omega^2 m/(EI)}$:
		\begin{itemize}
			\item For adjacent joints $i$ and $j$, the $(i,j)$ off-diagonal entry is $-C(\beta l_{ij})/A(\beta l_{ij})$, where $A(\beta l) = 1 - \cosh \beta l \cos \beta l$, $C(\beta l) = \sinh \beta l - \sin \beta l$, and $l_{ij}$ denotes the length of the beam segment connecting the joints $i$ and $j$;
			\item The $(i,i)$ diagonal entry is $\sum_j B(\beta l_{ij})/A(\beta l_{ij})$, where $B(\beta l) = \sinh \beta l \cos \beta l - \cosh \beta l \sin \beta l$, and the sum is over all beam segments around the joint $i$;
			\item The other entries are zero.
		\end{itemize}
		\item The Bloch-wave analysis is applicable to the above equation as usual for periodic structures.
	\end{itemize}
	The analytical functions as matrix elements come directly from the solution to the differential equation for beam vibration \citep{vibrationbeam}, and the frequency-dependent dynamical matrix $H$ is assembled using the balance conditions of the bending moments at the joints. The reader is referred to \cite{SUN2025105935} for further details.
	
	\section{Topological square grid-like frames}\label{secSquareFrames}
\begin{figure}[tb!] %1
		\centering
		\includegraphics[width=0.7\linewidth]{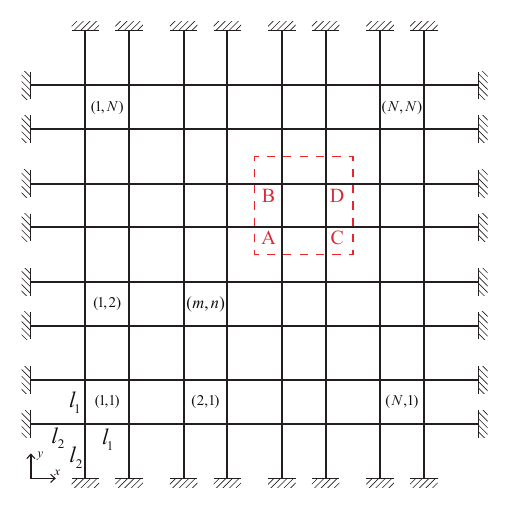}
		\caption{Topological square grid-like frame, with a unit cell outlined by red dashed lines.}
		\label{Figure_1}
	\end{figure}

	In this section, we consider the square grid-like frame with first-order and higher-order topological properties. As shown in Fig.~\ref{Figure_1}(a), the square frame consists of alternately arranged beam segments with lengths $l_1$ and $l_2$ along two orthogonal directions. All the exterior ends of the outermost beam segments are clamped. The structure contains a total of $N\times N$ unit cells, with $N$ unit cells along each edge, and each cell contains four rigid joints. The lengths of the intracell beam segments of a unit cell are $l_1$, and the lengths of the intercell beam segments are $l_2$. In the following, the existence of topological corner states and edge states in the spectrum of the square frame structure is investigated, and the frequencies of topological corner states, edge states and bulk states are given.
	
	First, the Bloch-wave analysis is carried out on the unit cell of the square frame, and the dynamical equation is obtained as
	\begin{equation}
		\label{eq5_9}
		H_{\mathrm{Bloch}}^{\mathrm{square}} \lvert \theta \rangle = 0,
	\end{equation}
	where $\lvert \theta \rangle$ represents a vector consisting of in-plane rotation angles of the four rigid joints in one unit cell, and the Bloch dynamical matrix is
	\begin{multline}
		\label{eq5_3}
		H_{\mathrm{Bloch}}^\mathrm{square}= \\
		\small
		\begin{bmatrix}
			\displaystyle 2\left[\frac{B(\beta l_1)}{A(\beta l_1)}+\frac{B(\beta l_2)}{A(\beta l_2)}\right] & \displaystyle -\frac{C(\beta l_1)}{A(\beta l_1)}-\frac{C(\beta l_2)}{A(\beta l_2)} \mathrm{e}^{-\mathrm{i}k_yL} &  \displaystyle -\frac{C(\beta l_1)}{A(\beta l_1)}-\frac{C(\beta l_2)}{A(\beta l_2)} \mathrm{e}^{-\mathrm{i}k_xL}  & 0\\[9pt]
			\displaystyle -\frac{C(\beta l_1)}{A(\beta l_1)}-\frac{C(\beta l_2)}{A(\beta l_2)} \mathrm{e}^{\mathrm{i}k_yL} & \displaystyle
			2\left[\frac{B(\beta l_1)}{A(\beta l_1)}+\frac{B(\beta l_2) }{A(\beta l_2) }\right]  & 0 &\displaystyle -\frac{C(\beta l_1)}{A(\beta l_1)}-\frac{C(\beta l_2)}{A(\beta l_2)} \mathrm{e}^{-\mathrm{i}k_xL}\\[9pt]
			\displaystyle -\frac{C(\beta l_1)}{A(\beta l_1)}-\frac{C(\beta l_2)}{A(\beta l_2)} \mathrm{e}^{\mathrm{i}k_xL} & 0 &\displaystyle 2\left[\frac{B(\beta l_1)}{A(\beta l_1)}+\frac{B(\beta l_2)}{A(\beta l_2)}\right]  &  \displaystyle -\frac{C(\beta l_1)}{A(\beta l_1)}-\frac{C(\beta l_2)}{A(\beta l_2)} \mathrm{e}^{-\mathrm{i}k_yL} \\[9pt] 0 & 	\displaystyle -\frac{C(\beta l_1)}{A(\beta l_1)}-\frac{C(\beta l_2)}{A(\beta l_2)} \mathrm{e}^{\mathrm{i}k_xL} & \displaystyle -\frac{C(\beta l_1)}{A(\beta l_1)}-\frac{C(\beta l_2)}{A(\beta l_2)} \mathrm{e}^{\mathrm{i}k_yL} & \displaystyle 2\left[\frac{B(\beta l_1)}{A(\beta l_1)}+\frac{B(\beta l_2)}{A(\beta l_2)}\right]
		\end{bmatrix},
	\end{multline}
	where $k_x$ and $k_y$ are the wavenumbers in the two orthogonal directions, $L = l_1 + l_2$, and the definitions of functions $A$, $B$ and $C$ are
	\begin{align}
		\label{eqa2}
		A(\beta l)&\equiv 1-\cosh\beta l\cos\beta l, \\
		\label{eqa3}
		B(\beta l)&\equiv \sinh\beta l\cos\beta l-\cosh\beta l\sin\beta l, \\
		\label{eqa4}
		C(\beta l)&\equiv \sinh\beta l-\sin\beta l.
	\end{align}
	Note that Eq.~\eqref{eq5_9} amounts to saying that $H_{\mathrm{Bloch}}^{\mathrm{square}}$ must have a zero eigenvalue, for which $\lvert \theta \rangle$ is the corresponding eigenvector. The dynamical matrix can be written in the form of a Kronecker sum:
	\begin{equation}
		\label{eq5_4}
		H_{\mathrm{Bloch}}^\mathrm{square}=H_{\mathrm{Bloch}}^\mathrm{beam}(k_x) \otimes I + I \otimes H_{\mathrm{Bloch}}^\mathrm{beam}(k_y),
	\end{equation}
	where $H_{\mathrm{Bloch}}^\mathrm{beam}$ is the Bloch dynamical matrix of the periodic 1D \emph{continuous beam structure} with alternating spans \citep{SUN2025105935}, whose expression is
	\begin{equation}
		\label{eq24}
		H_{\mathrm{Bloch}}^\mathrm{beam} (k) =
		\begin{bmatrix}
			\displaystyle \frac{B(\beta l_1)}{A(\beta l_1)}+\frac{B(\beta l_2)}{A(\beta l_2)} & \displaystyle -\frac{C(\beta l_1)}{A(\beta l_1)}-\frac{C(\beta l_2)}{A(\beta l_2)} \exp(-\mathrm{i}kL)\\[10pt]
			\displaystyle -\frac{C(\beta l_1)}{A(\beta l_1)}-\frac{C(\beta l_2)}{A(\beta l_2)} \exp(\mathrm{i}kL) & \displaystyle
			\frac{B(\beta l_1)}{A(\beta l_1)}+\frac{B(\beta l_2) }{A(\beta l_2) }
		\end{bmatrix},
	\end{equation}
	where $k$ is the wavenumber.
	
	For a finite-sized square frame with boundaries, its dynamical matrix can be written in an analogous manner:
	\begin{equation}
		\label{eq5_5}
		H^\mathrm{square}=H^\mathrm{beam} \otimes I + I \otimes H^\mathrm{beam},
	\end{equation}
	where $H^\mathrm{beam}$ is the dynamical matrix of a finite continuous beam \citep{SUN2025105935}.
	
	In order to obtain the eigenvalue spectrum according to the dynamical matrix, a few key facts should be noted first. The Kronecker sum $H_{xy} = H_x \otimes I + I \otimes H_y$ of matrices $H_x$ and $H_y$ has the following property: If $\lvert\theta_x\rangle$ and $\lambda_x$ are an eigenvector--eigenvalue pair of $H_x$, $\lvert\theta_y\rangle$ and $\lambda_y$ are an eigenvector--eigenvalue pair of $H_y$, then
	\begin{equation}
		\label{kro}
		\begin{split}
			H_{xy} \bigl( \lvert\theta_x\rangle \otimes \lvert\theta_y\rangle \bigr) &= (H_x \otimes I + I \otimes H_y) \bigl( \lvert\theta_x\rangle \otimes \lvert\theta_y\rangle \bigr) \\
			&= ( H_x \lvert\theta_x\rangle ) \otimes \lvert\theta_y\rangle + \lvert\theta_x\rangle \otimes ( H_y \lvert\theta_y\rangle ) \\
			&= \lambda_x \lvert\theta_x\rangle \otimes \lvert\theta_y\rangle + \lvert\theta_x\rangle \otimes \lambda_y \lvert\theta_y\rangle \\
			&= (\lambda_x + \lambda_y) \bigl( \lvert\theta_x\rangle \otimes \lvert\theta_y\rangle \bigr),
		\end{split}
	\end{equation}
	that is,
	\begin{itemize}
		\item The eigenvectors of the Kronecker sum $H_{xy}$ are the Kronecker products $\lvert\theta_x\rangle \otimes \lvert\theta_y\rangle$ of the eigenvectors of the original matrices $H_x$ and $H_y$;
		\item The corresponding eigenvalues are simply the sums $\lambda_x + \lambda_y$ of the eigenvalues of original matrices.
	\end{itemize}
	Therefore, the eigenvalue spectrum of the dynamical matrix $H^\mathrm{square}$ of the square grid-like frame can be obtained by performing pairwise addition of the eigenvalues of the dynamical matrices $H^\mathrm{beam}$ of two identical 1D continuous beam structures.
	On the other hand, for any given $\beta$, the form of the dynamical matrix $H^{\mathrm{beam}}$ (or its Bloch counterpart, presented in Eq.~\eqref{eq24}) for a 1D continuous beam is completely identical to the Hamiltonian of a Su--Schrieffer--Heeger (SSH) chain \citep{SSH1979} (up to subtraction of a scalar matrix), with $-C(\beta l_1)/A(\beta l_1)$ and $-C(\beta l_2)/A(\beta l_2)$ in the role of intracell and intercell hopping strengths between sites, respectively. This analogy with a well-studied topological model can be used as an intermediate step to deduce the band structure and topological properties of the continuous beam structure \citep{SUN2025105935}, and hence the square grid-like frame. In the following, we derive the frequency ranges of different modes (i.e., corner, edge, and bulk states) in the spectrum, and analyze the distribution patterns of their frequencies with an example structure.
	
	\subsection{Higher-order topological corner states in square frames}\label{secSquareCorner}
	The frequencies $\beta_t$ of topological corner states in a finite square grid-like frame with fixed-end boundaries must satisfy \citep{SUN2025105935}
	\begin{equation}
		\label{eq5_1}
		2 \left[\frac{B(\beta_t l_1)}{A(\beta_t l_1)}+\frac{B(\beta_t l_2)}{A(\beta_t l_2)}\right] = 0.
	\end{equation}
	In each positive interval $(\beta_0^{(n-1)},\beta_0^{(n)})$ (where $\beta_0^{(n)}$ is defined as the $n$-th positive root of $A(\beta l_{1}) A(\beta l_{2}) = 0$, and $\beta_0^{(0)}=0$), there exists one $\beta_t$ satisfying Eq.~\eqref{eq5_1}, denoted by $\beta_t^{(n)}$ (a detailed proof is given by \citet{SUN2025105935}); notably, there is an infinite number of such $\beta_t^{(n)}$ in theory. For a specific $\beta_t^{(n)}$, if and only if the off-diagonal elements of matrix $H_{\mathrm{Bloch}}^{\mathrm{square}}$ satisfy the topological nontriviality condition \citep{SUN2025105935}
	\begin{equation}
		\label{eq5_2}
		\left \lvert \frac{C(\beta_t^{(n)} l_1)}{A(\beta_t^{(n)} l_1)} \right \rvert < \left \lvert\frac{C(\beta_t^{(n)} l_2)}{A(\beta_t^{(n)} l_2)} \right \rvert
	\end{equation}
	do topological corner states exist at frequency $\beta_t^{(n)}$. %Condition~\eqref{eq5_2} is due to the existence condition of edge eigenstates $\lvert\theta_x\rangle$ and $\lvert\theta_y\rangle$ of the 1D continuous beam structure \citep{SUN2025105935}, which has a form analogous to the topological nontriviality condition of the SSH chain.
	
	\begin{figure}[tbp] %2
		\centering
		\includegraphics[width=0.95\linewidth]{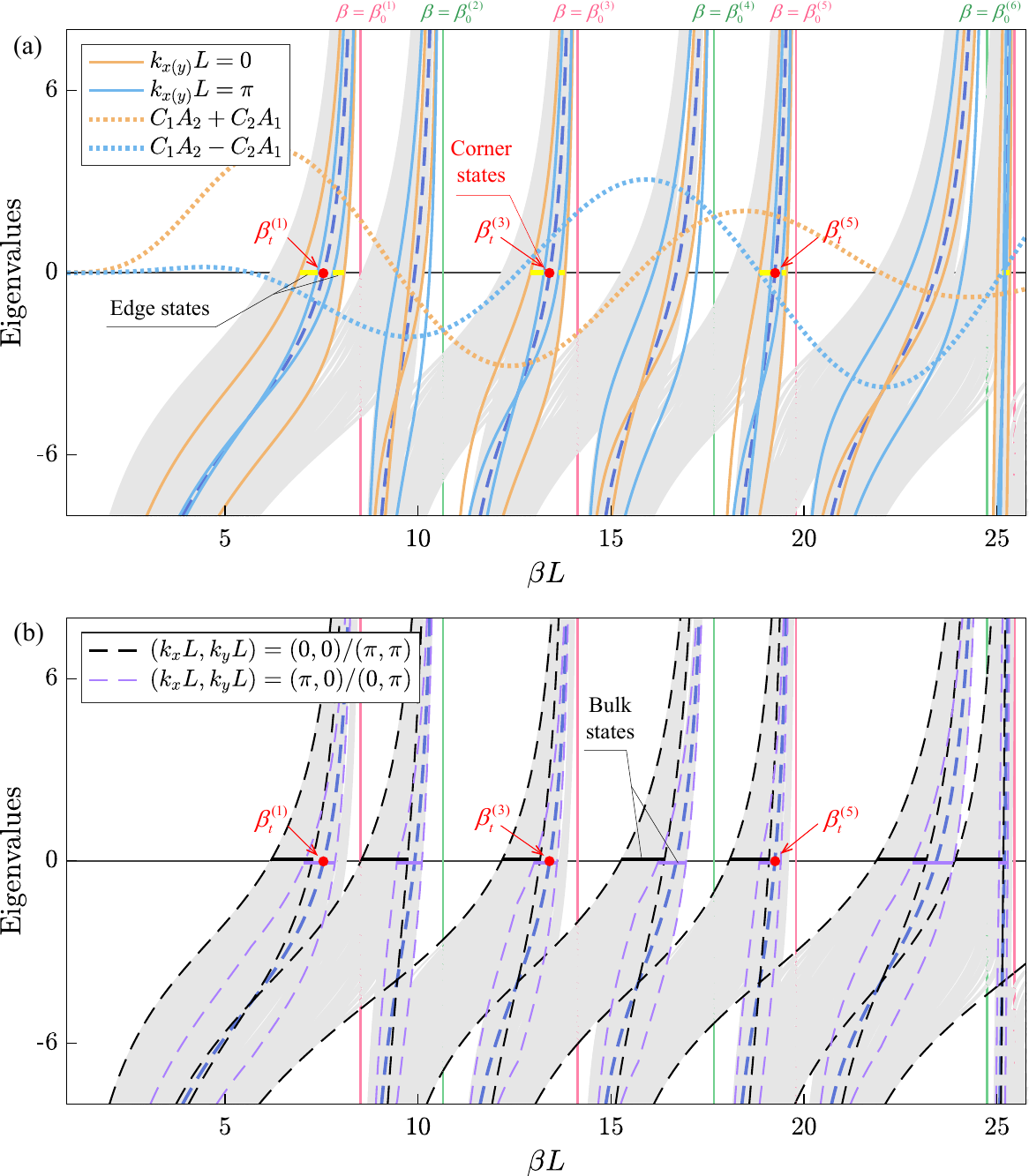}
		\caption{Eigenvalue spectrum of a topological square grid-like frame, with lengths $l_1=40\,\mathrm{mm}$, $l_2=50\,\mathrm{mm}$. The gray curves represent the eigenvalues of the dynamical matrix $H^{\mathrm{square}}$ in Eq.~\eqref{eq5_5} of the finite-sized frame, and all the intersections of the gray curves and the black horizontal line $\lambda=0$ constitute the frequency spectrum of the frame. Green (pink) vertical lines represent the roots of the equation $A(\beta_0 l_{1(2)})=0$. The intersections of dark blue dashed curves with $\lambda=0$ pointed by red arrows indicate the frequencies of topological corner states. (a) Edge states are identified by yellow line segments, where the values of the functions $C(\beta l_1)A(\beta l_2)+C(\beta l_2)A(\beta l_1)$ (orange dotted curve) and $C(\beta l_1)A(\beta l_2)-C(\beta l_2)A(\beta l_1)$ (blue dotted curve) have opposite signs. To ensure that the functions corresponding to the orange and blue dotted curves are clearly plotted in the window, the ordinates of both have been scaled appropriately (the scaling factor is proportional to $\exp[-\beta(l_1+l_2)]$, which does not affect the sign of the function). (b) Bulk bands correspond to regions on the $\lambda=0$ line bounded by adjacent black dashed curves or adjacent purple dashed curves.}
		\label{Figure_2}
	\end{figure}
	
	In the following, we show that the higher-order topological corner states of the structure are immensely overlapped with the bulk bands in frequency, and identify the corner states with the use of Eq.~\eqref{eq5_1}. As an example, we take the square grid-like frame with parameters $l_1=40~\mathrm{mm}$, $l_2=50~\mathrm{mm}$, and $N=8$. The eigenvalue spectrum of the dynamical matrix $H^{\mathrm{square}}$ in Eq.~\eqref{eq5_5} of the square frame is shown by gray curves in Fig.~\ref{Figure_2}. All the intersections of the gray curves and the horizontal line $\lambda=0$ (where $\lambda$ is the ordinate), that is, all $\beta$ at which an eigenvalue of the matrix $H^{\mathrm{square}} (\beta)$ is zero, constitute the entire spectrum of the square frame. The horizontal positions of pink vertical lines ($A(\beta_0 l_2)=0$) and green vertical lines ($A(\beta_0 l_1)=0$) represent the frequencies $\beta=\beta_0^{(n)}$, and between every two adjacent vertical lines exists a dark blue dashed curve $\lambda=2\left[\frac{B(\beta l_1)}{A(\beta l_1)}+\frac{B(\beta l_2)}{A(\beta l_2)}\right]$, which is the left-hand side of Eq.~\eqref{eq5_1}; $\beta_t^{(n)}$ is represented by the intersections of dark blue dashed curves and the line $\lambda=0$. 
	 
	 Now we note that it is impractical to identify the higher-order topological corner states by directly observing the eigenvalue spectrum calculated from Eq.~\eqref{eq5_5} in Fig.~\ref{Figure_2}, as one cannot tell whether a gray curve in the vicinity of the dark blue dashed line represents a bulk state or a corner state; actually, all the topological corner states must lie within the frequency range of bulk states (demonstrated in Section~\ref{secSquareBulk} below). In Fig.~\ref{Figure_2}, the existence of corner states is judged by the signs of the functions
	\begin{equation}
		C(\beta l_1)A(\beta l_2) \pm C(\beta l_2)A(\beta l_1)
	\end{equation}
	represented by the orange and blue dotted curves; for a specific frequency $\beta_t^{(n)}$, if the signs of the ordinates of blue and orange dotted curves are opposite, the topological corner states exist at frequency $\beta_t^{(n)}$; if the signs of the ordinates of blue and orange dotted curves are the same, the topological corner states do not exist at frequency $\beta_t^{(n)}$. This method of judgment is exactly equivalent to the criterion~\eqref{eq5_2}. As a result, in the range $(0,\beta_0^{(6)})$ in Fig.~\ref{Figure_2}, topological corner states only exist at frequencies $\beta_t^{(1)}$, $\beta_t^{(3)}$ and $\beta_t^{(5)}$. Although the corner states at these frequencies are degenerate with the bulk states, due to which usual numerical solving approaches may give hybrid modes as results, we can always extract the  corner modes (whose displacements are localized at the corners) from multiple degenerate eigenmodes at a single frequency (that is, displacements are localized at the corners), by means of singular value decomposition\footnote{Essentially, multiple modes with the same (i.e., degenerate) eigenvalue can be arbitrarily linearly superposed to produce another eigenmode; singular value decomposition only serves to determine the appropriate superposition coefficients.}. 
	
\begin{figure}[tb!] %3
		\centering
		\includegraphics[width=0.9\linewidth]{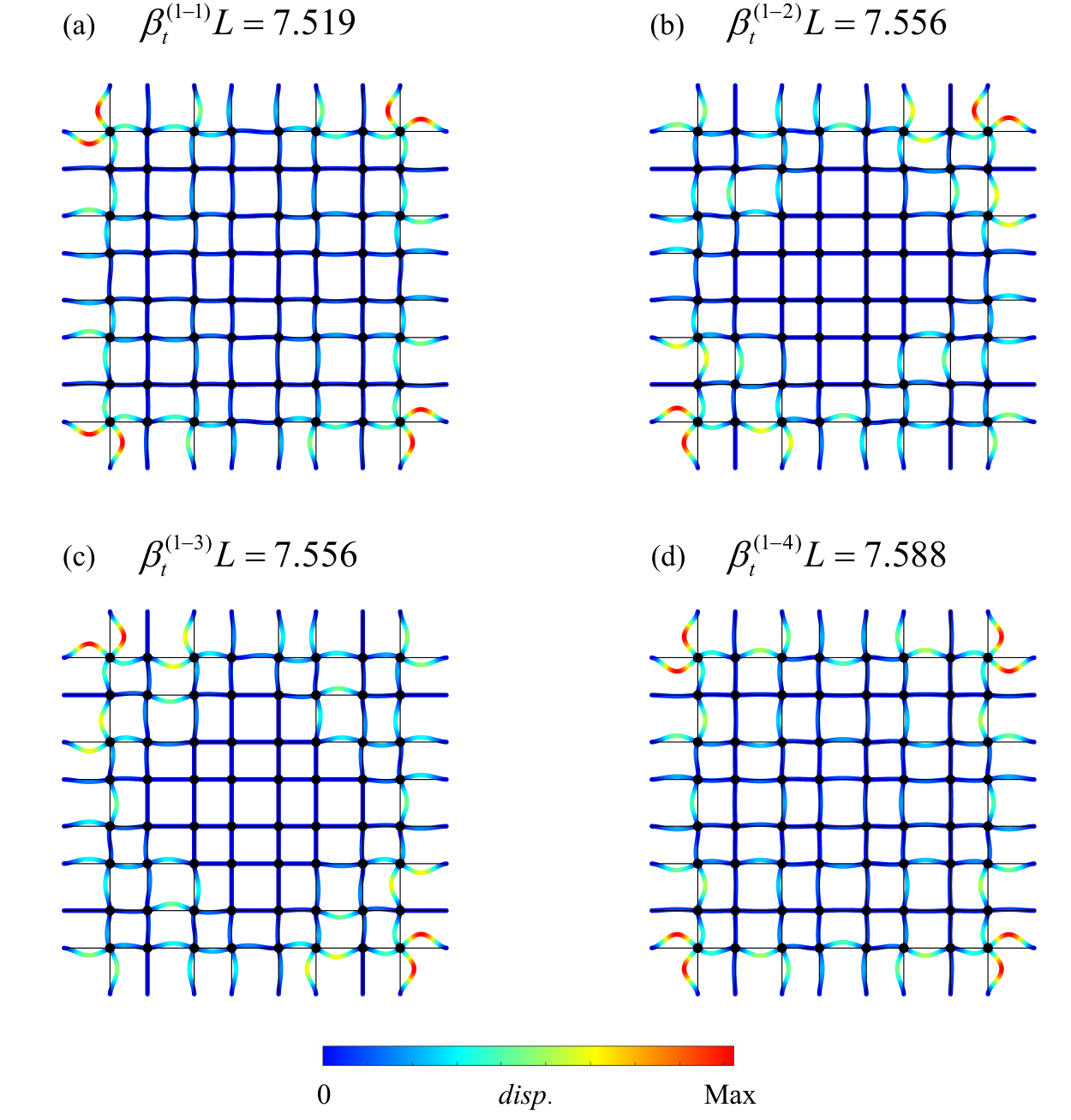}
		\caption{Topological corner modes of square grid-like frame, where the modes are quadruply degenerate at frequency $\beta_{t}^{(1)}$, and are localized at the corners.}
		\label{Figure_3}
	\end{figure}

	The mode shapes of topological corner states at frequency $\beta_t^{(1)}$ extracted from the bulk states are shown in Fig.~\ref{Figure_3}, where the bending deformations of the beam segments are localized at the corners of the square frame. The eigenspectrum in Fig.~\ref{Figure_2} and eigenvectors of the matrix $H^\mathrm{square}$ (whose elements are analytical expressions) are calculated with the software MATLAB, and then we plot the beam deflections in Fig.~\ref{Figure_3} according to the eigenvector $\lvert\theta\rangle$ of rotation angles at joints. Since the structure has $C_4$ symmetry, the modes in Fig.~\ref{Figure_3}(a) and Fig.~\ref{Figure_3}(d) exhibit $C_4$ symmetry and $C_4$ antisymmetry, respectively, and are localized simultaneously at the four corners; meanwhile, linear superpositions of the two degenerate modes in Fig.~\ref{Figure_3}(b)--(c) can also yield two modes that are eigenmodes of the $C_4$-rotation operator (with eigenvalues $\pm \mathrm{i}$).
	
	\subsection{Edge states in square frames}\label{secSquareEdge}
	Due to the aforementioned property of the Kronecker sum, we recognize that for a square frame (whose dynamical matrix is given in Eq.~\eqref{eq5_5}), every eigenvalue $\lambda^\mathrm{square}$ of the dynamical matrix $H^\mathrm{square}$ of the square grid-like frame can be obtained by adding two eigenvalues $\lambda^\mathrm{beam}$ of the dynamical matrices $H^\mathrm{beam}$ of two continuous beams with the same structural parameters. Now, as an edge state of the 2D structure must come from the tensor product of two 1D states---an \emph{edge} state along one direction and a \emph{bulk} state along another direction, the eigenvalue $\lambda^\mathrm{square}_\mathrm{edge}$ of matrix $H^\mathrm{square}$ corresponding to an edge state of the square frame must equal the sum of the eigenvalue $\lambda^\mathrm{beam}_\mathrm{edge}$ of matrix $H^\mathrm{beam}$ corresponding to an edge state, and the eigenvalue $\lambda^\mathrm{beam}_\mathrm{bulk}$ of matrix $H^\mathrm{beam}$ corresponding to a bulk state. Through the analogy in matrix form
	with the Hamiltonian of the SSH chain (whose topological edge states lie at zero energy), we have
	\begin{equation}
		\label{eq5_6}
		\lambda^\mathrm{beam}_\mathrm{edge}=\left[\frac{B(\beta l_1)}{A(\beta l_1)}+\frac{B(\beta l_2)}{A(\beta l_2)}\right]+0.
	\end{equation}
	On the other hand, it follows from Eq.~\eqref{eq24} that
	\begin{equation}
		\label{eq5_7}
		\lambda^\mathrm{beam}_\mathrm{bulk}=\left[\frac{B(\beta l_1)}{A(\beta l_1)}+\frac{B(\beta l_2)}{A(\beta l_2)}\right]\pm \left\lvert\frac{C(\beta l_1)}{A(\beta l_1)} + \frac{C(\beta l_2)}{A(\beta l_2)} \exp(\mathrm{i}k_{x(y)}L)\right\rvert,
	\end{equation}
	where $k_{x(y)}L\in(-\pi,\pi]$. Thus, we obtain
	\begin{equation}
		\label{eq5_8}
		\lambda^\mathrm{square}_\mathrm{edge} = \lambda^\mathrm{beam}_\mathrm{edge} + \lambda^\mathrm{beam}_\mathrm{bulk} = 2\left[\frac{B(\beta l_1)}{A(\beta l_1)}+\frac{B(\beta l_2)}{A(\beta l_2)}\right]\pm \left\lvert\frac{C(\beta l_1)}{A(\beta l_1)} + \frac{C(\beta l_2)}{A(\beta l_2)} \exp(\mathrm{i}k_{x(y)}L)\right\rvert.
	\end{equation}
	In order for the dynamical equation $H^{\mathrm{square}} \lvert \theta \rangle = 0$ of the square frame to have an edge-state solution, it holds that $\lambda^\mathrm{square}_\mathrm{edge}=0$, that is, the frequency $\beta$ of the edge state should satisfy \citep{SUN2025105935}
	\begin{equation}
		\label{eq5_10}
		2\left[\frac{B(\beta l_1)}{A(\beta l_1)}+\frac{B(\beta l_2)}{A(\beta l_2)}\right]=\pm \left\lvert\frac{C(\beta l_1)}{A(\beta l_1)} + \frac{C(\beta l_2)}{A(\beta l_2)} \exp(\mathrm{i}k_{x(y)}L)\right\rvert.
	\end{equation}
	Moreover, the topological nontriviality condition should apply: for $\beta$ satisfying Eq.~\eqref{eq5_10}, if and only if
	\begin{equation}
		\label{eq5_10a}
		\left \lvert \frac{C(\beta l_1)}{A(\beta l_1)} \right \rvert < \left \lvert\frac{C(\beta l_2)}{A(\beta l_2)} \right \rvert,
	\end{equation}
	edge states exist at frequencies $\beta$.
	
	\begin{figure}[tb!] %4
		\centering
		\includegraphics[width=0.9\linewidth]{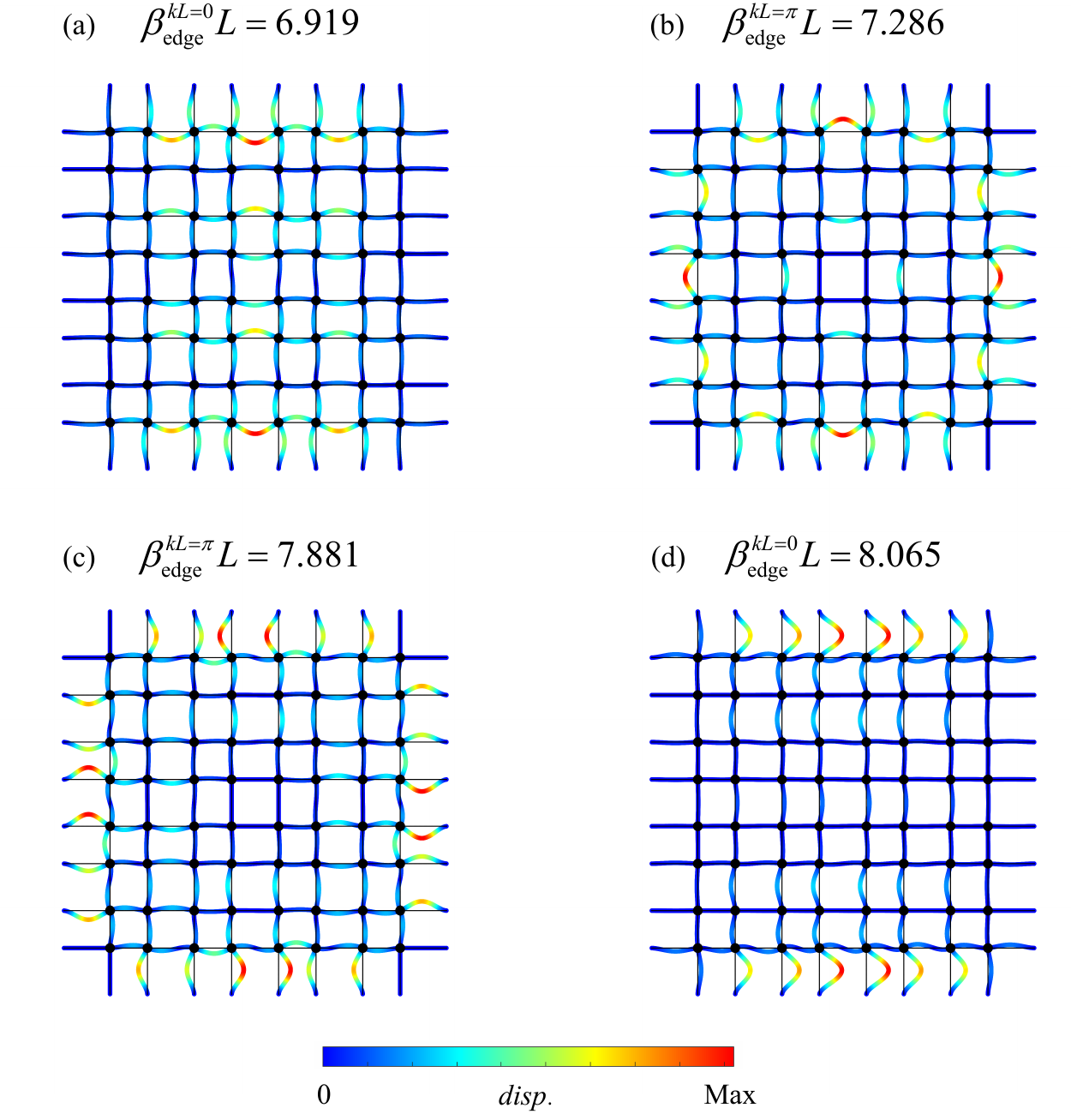}
		\caption{Edge modes of the square grid-like frame. (a)--(b) Edge states at $k_{x(y)}L=0$, $k_{x(y)}L=\pi$ in the interval $(\beta_0^{(0)},\beta_t^{(1)})$ in Fig.~\ref{Figure_2}. (c)--(d) Edge states at $k_{x(y)}L=\pi$, $k_{x(y)}L=0$ in the interval $(\beta_t^{(1)},\beta_0^{(1)})$ in Fig.~\ref{Figure_2}.}
		\label{Figure_4}
	\end{figure}
	
	We prove the existence of solutions for Eq.~\eqref{eq5_10} below. As we will see, there exists a set of continuous $\beta$ satisfying Eq.~\eqref{eq5_10} in every interval $(\beta_0^{(n-1)},\beta_t^{(n)})$ and $(\beta_t^{(n)},\beta_0^{(n)})$. Meanwhile, when $\left \lvert \frac{C(\beta_t l_1)}{A(\beta_t l_1)} \right \rvert \neq\left \lvert\frac{C(\beta_t l_2)}{A(\beta_t l_2)} \right \rvert$, topological corner states must lie within bandgaps of edge states, that is, the frequencies of corner states are not inside the frequency range of edge states. To this end, we let
	%	\begin{equation}
		%		\label{eq5_11}
		\[
		f(\beta)=\left\lvert\frac{C(\beta l_1)}{A(\beta l_1)} + \frac{C(\beta l_2)}{A(\beta l_2)} \exp(\mathrm{i}k_{x(y)}L)\right\rvert^2 - 4\left[\frac{B(\beta l_1)}{A(\beta l_1)} + \frac{B(\beta l_2)}{A(\beta l_2)}\right]^2.
		\]
		%	\end{equation}
	The solution set of $f(\beta)=0$ is the frequency range where edge states may exist (see Eq.~\eqref{eq5_10}). Now consider $f(\beta_t^{(n)})$. First note that Eq.~\eqref{eq5_1} holds at the frequencies $\beta_t$ of the corner states, which implies that the second term of $f(\beta_t^{(n)})$ is zero. Next, when $\left \lvert \frac{C(\beta_t l_1)}{A(\beta_t l_1)} \right \rvert \neq\left \lvert\frac{C(\beta_t l_2)}{A(\beta_t l_2)} \right \rvert$, we have
	\[
	\left\lvert\frac{C(\beta_t l_1)}{A(\beta_t l_1)} + \frac{C(\beta_t l_2)}{A(\beta_t l_2)} \exp(\mathrm{i}k_{x(y)}L)\right\rvert \ge \Biggl\lvert \left\lvert \frac{C(\beta_t l_1)}{A(\beta_t l_1)} \right\rvert - \left\lvert \frac{C(\beta_t l_2)}{A(\beta_t l_2)} \right\rvert \Biggr\rvert > 0.
	\]
	Therefore, for $\beta=\beta_t^{(n)}$,
	\[ f(\beta_t^{(n)}) > 0, \]
	and Eq.~\eqref{eq5_10} never holds. It is concluded that as long as $\left \lvert \frac{C(\beta_t l_1)}{A(\beta_t l_1)} \right \rvert \neq\left \lvert\frac{C(\beta_t l_2)}{A(\beta_t l_2)} \right \rvert$, the frequencies of corner states do not coincide with those of edge states. On the other hand, when $\beta\rightarrow\beta_0^{(n)}$ (i.e., $A(\beta l_{1})\rightarrow 0^\pm$ or $A(\beta l_{2})\rightarrow 0^\pm$), in the light of the identity $C^2(\beta l)-B^2(\beta l)=-2\sinh(\beta l)\sin(\beta l)\cdot A(\beta l)$, it holds that
	\[
	f(\beta)=-3\left[\frac{B(\beta l_{1(2)})}{A(\beta l_{1(2)})}\right]^2+O[(\beta-\beta^{(n)}_0)^{-1}]\rightarrow -\infty;
	\]
	we note that the order of the first term is $(\beta - \beta_0^{(n)})^{-2}$, and thus $f(\beta)$ tends to negative infinity on both sides of $\beta_0^{(n)}$. In terms of this result and the condition $f(\beta_t^{(n)})>0$, there must exist a set of continuous $\beta$ satisfying Eq.~\eqref{eq5_10} in each interval $(\beta_0^{(n-1)},\beta_t^{(n)})$ and $(\beta_t^{(n)},\beta_0^{(n)})$. That is to say, the frequency $\beta_t^{(n)}$ of the corner states lies within the bandgaps of candidate edge states\footnote{The solutions to Eq.~\eqref{eq5_10} correspond to the frequencies of candidate edge states.}.
	
	The above conclusions are verified by the example structure, whose eigenvalue spectrum is shown in Fig.~\ref{Figure_2}(a). For any specific $\beta$, when the sign on the right-hand side of Eq.~\eqref{eq5_8} is fixed and taken to be either plus or minus, the eigenvalue $\lambda^\mathrm{square}_\mathrm{edge}$ is monotonic with respect to $k_{x(y)}L$ in the range $k_{x(y)}L\in[0,\pi]$, and the lower and upper bounds of  $\lambda^\mathrm{square}_\mathrm{edge}$ are reached at $k_{x(y)}L=0$ or $\pi$, represented by orange and blue solid curves as shown in Fig.~\ref{Figure_2}(a). There exist a blue and an orange solid curve between each dark blue dashed curve and the neighboring pink or green vertical line. The frequencies $\beta$ that satisfy Eq.~\eqref{eq5_10} correspond to the intervals bounded by intersections of adjacent blue and orange curves with axis $\lambda=0$, within the ranges $(\beta_0^{(n-1)},\beta_t^{(n)})$ and $(\beta_t^{(n)},\beta_0^{(n)})$.
	
	Now we show that edge states only exist at $\beta$ which satisfy both Eqs.~\eqref{eq5_10} and \eqref{eq5_10a}. We note that only in part of the intervals defined by Eq.~\eqref{eq5_10} are edge states present, indicated by yellow line segments in Fig.~\ref{Figure_2}(a), where the two functions $[C(\beta l_1)A(\beta l_2) \pm C(\beta l_2)A(\beta l_1)]$ corresponding to the blue and orange dotted curves take opposite signs, rendering the condition~\eqref{eq5_10a} true. The edge modes of the square grid-like frame in intervals $(0,\beta_t^{(1)})$ and $(\beta_t^{(1)},\beta_0^{(1)})$ are shown in Fig.~\ref{Figure_4}, where the deformations of beams are localized at the edges of square frames. The above conclusions are verified.
	
	\subsection{Bulk states in square frames}\label{secSquareBulk}
\begin{figure}[tb!] %5
		\centering
		\includegraphics[width=\linewidth]{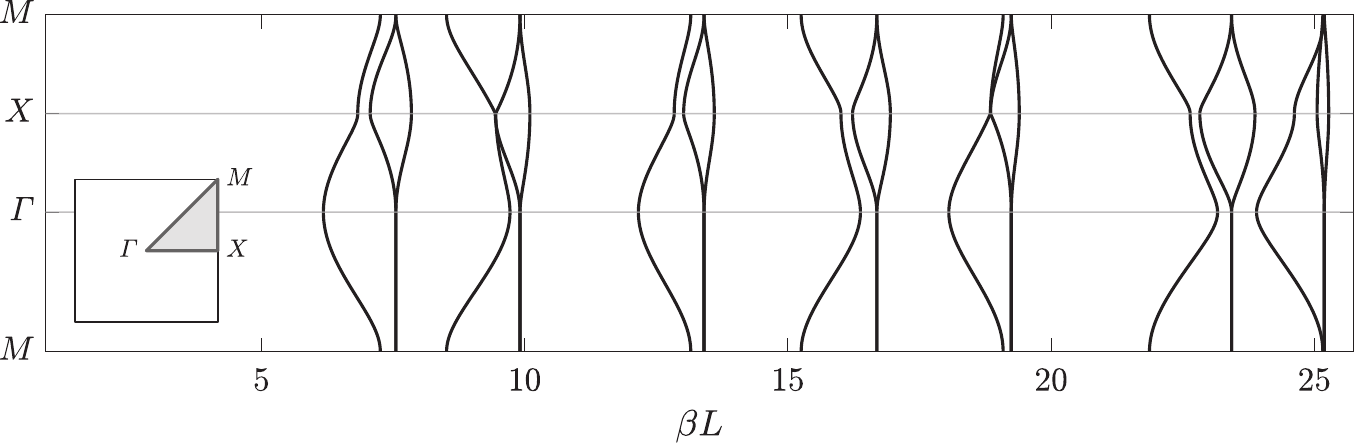}
		\caption{Band structure of the square frame with $(l_1, l_2) = (40\,\mathrm{mm}, 50\,\mathrm{mm})$ or $(50\,\mathrm{mm}, 40\,\mathrm{mm})$. Inset shows the first Brillouin zone.}
		\label{Figure_5}
	\end{figure}

	Due to the property of the Kronecker sum, the eigenvalue $\lambda^\mathrm{square}_\mathrm{bulk}$ of matrix $H^\mathrm{square}$ corresponding to any bulk state of the square grid-like frame equals the sum of two eigenvalues $\lambda^\mathrm{beam}_\mathrm{bulk}$ of matrix $H^\mathrm{beam}$ corresponding to bulk states of the continuous beam in the two orthogonal directions. It follows from Eq.~\eqref{eq5_7} that
	\begin{equation}
		\label{eq5_12}
		\lambda^\mathrm{square}_\mathrm{bulk} = 2\left[\frac{B(\beta l_1)}{A(\beta l_1)}+\frac{B(\beta l_2)}{A(\beta l_2)}\right] \pm \left\lvert\frac{C(\beta l_1)}{A(\beta l_1)} + \frac{C(\beta l_2)}{A(\beta l_2)} \exp(\mathrm{i}k_{x}L)\right\rvert \pm \left\lvert\frac{C(\beta l_1)}{A(\beta l_1)} + \frac{C(\beta l_2)}{A(\beta l_2)} \exp(\mathrm{i}k_{y}L)\right\rvert.
	\end{equation}
	For a specific $\beta$, the set of eigenvalues $\lambda^\mathrm{square}_\mathrm{bulk}$ is the union of three clusters, denoted as top, middle and bottom, given by the expressions
	{\small
		\begin{align}
			\label{eq5_13}
			\lambda^\mathrm{square}_\mathrm{bulk(top)} &= 2\left[\frac{B(\beta l_1)}{A(\beta l_1)}+\frac{B(\beta l_2)}{A(\beta l_2)}\right] + \left\lvert\frac{C(\beta l_1)}{A(\beta l_1)} + \frac{C(\beta l_2)}{A(\beta l_2)} \exp(\mathrm{i}k_{x}L)\right\rvert + \left\lvert\frac{C(\beta l_1)}{A(\beta l_1)} + \frac{C(\beta l_2)}{A(\beta l_2)} \exp(\mathrm{i}k_{y}L)\right\rvert, \\
			\label{eq5_14}
			\lambda^\mathrm{square}_\mathrm{bulk(middle)} &= 2\left[\frac{B(\beta l_1)}{A(\beta l_1)}+\frac{B(\beta l_2)}{A(\beta l_2)}\right] + \left\lvert\frac{C(\beta l_1)}{A(\beta l_1)} + \frac{C(\beta l_2)}{A(\beta l_2)} \exp(\mathrm{i}k_{x(y)}L)\right\rvert - \left\lvert\frac{C(\beta l_1)}{A(\beta l_1)} + \frac{C(\beta l_2)}{A(\beta l_2)} \exp(\mathrm{i}k_{y(x)}L)\right\rvert, \\
			\label{eq5_15}
			\lambda^\mathrm{square}_\mathrm{bulk(bottom)} &= 2\left[\frac{B(\beta l_1)}{A(\beta l_1)}+\frac{B(\beta l_2)}{A(\beta l_2)}\right] - \left\lvert\frac{C(\beta l_1)}{A(\beta l_1)} + \frac{C(\beta l_2)}{A(\beta l_2)} \exp(\mathrm{i}k_{x}L)\right\rvert - \left\lvert\frac{C(\beta l_1)}{A(\beta l_1)} + \frac{C(\beta l_2)}{A(\beta l_2)} \exp(\mathrm{i}k_{y}L)\right\rvert.
		\end{align}
	}In order for the dynamical equation \eqref{eq5_9} of square frames to have nontrivial solutions, $\lambda^\mathrm{square}_\mathrm{bulk}=0$ needs to be satisfied, which corresponds to the bulk states of square grid-like frames. The dispersion diagram of the bulk bands for a square frame with $(l_1, l_2) = (40\,\mathrm{mm}, 50\,\mathrm{mm})$ is presented in Fig.~\ref{Figure_5}.

	As shown in Fig.~\ref{Figure_2}(b), the frequency ranges of bulk states satisfying $\lambda^\mathrm{square}_\mathrm{bulk(top)}=0$ or $\lambda^\mathrm{square}_\mathrm{bulk(bottom)}=0$ correspond to the set of points on the horizontal axis that lie between two neighboring black dashed curves, and the frequency ranges of bulk states that satisfy $\lambda^\mathrm{square}_\mathrm{bulk(middle)}=0$ correspond to the set of points between two neighboring purple dashed curves. It is interesting to note that the set of eigenvalues $\lambda^{\mathrm{square}}_{\mathrm{bulk(top)}}$ corresponding to Eq.~\eqref{eq5_13} has exactly the same range as given by the expression
	\begin{equation}	\label{eq5_13add}
		2\left[\frac{B(\beta l_1)}{A(\beta l_1)}+\frac{B(\beta l_2)}{A(\beta l_2)}\right] + 2\left\lvert\frac{C(\beta l_1)}{A(\beta l_1)} + \frac{C(\beta l_2)}{A(\beta l_2)} \exp(\mathrm{i}kL)\right\rvert.
	\end{equation}
	The set of eigenvalues $\lambda^{\mathrm{square}}_{\mathrm{bulk(bottom)}}$ corresponding to Eq.~\eqref{eq5_15} has exactly the same range as given by the expression
	\begin{equation}	\label{eq5_14add}
		2\left[\frac{B(\beta l_1)}{A(\beta l_1)}+\frac{B(\beta l_2)}{A(\beta l_2)}\right] - 2\left\lvert\frac{C(\beta l_1)}{A(\beta l_1)} + \frac{C(\beta l_2)}{A(\beta l_2)} \exp(\mathrm{i}kL)\right\rvert.
	\end{equation}
	By comparing the expressions~\eqref{eq5_13add} and \eqref{eq5_14add} with Eq.~\eqref{eq5_7}, it is concluded that part of the bulk bands of the square frame, that is, the bulk frequency ranges pertaining to the ``top'' and ``bottom'' clusters ($\lambda^\mathrm{square}_\mathrm{bulk(top)} = 0$ or $\lambda^\mathrm{square}_\mathrm{bulk(bottom)} = 0$), is identical to the frequency ranges of bulk bands of the continuous beam structure \citep{SUN2025105935}. Meanwhile, the frequency ranges of bulk states satisfying $\lambda^\mathrm{square}_\mathrm{bulk(middle)}=0$ (the ``middle'' cluster) must contain all the frequencies of corner states $\beta_t$, since the range of values of Eq.~\eqref{eq5_14} for different $(k_x, k_y)$ always includes the value $\lambda^\mathrm{square}_\mathrm{bulk(middle)}=2\left[\frac{B(\beta l_1)}{A(\beta l_1)}+\frac{B(\beta l_2)}{A(\beta l_2)}\right]+0=\lambda^\mathrm{square}_\mathrm{corner}$. This conclusion is also graphically shown from Fig.~\ref{Figure_2}(b), where the dark blue dashed curves are always immersed in the shaded regions between adjacent purple dashed curves.
	
	\subsection{Remarks on the Kronecker-sum approach}
	The above solution procedure in this section intensively makes use of the mathematical property of the Kronecker sum, which provides an efficient way to decompose the 2D problem into simpler ones of the 1D constituents. Such technique is actually not limited to solving the square frame, but also applies to frames with rectangular or even parallelogram geometries. We note that for a dynamical matrix $H_{xy}$ to admit a Kronecker-sum form $H_{xy}=H_x\otimes I + I\otimes H_y$, it is required that all rows of the frame structure be identical (and so should all columns); the boundaries of the whole frame structure must be parallel to the 1D constituent rows and columns; furthermore, each joint can only be connected to joints within the same row and the same column, for which reason the application of this approach to the kagome lattice would not be possible. The solution of the kagome frame employs a distinct approach, detailed in Section~\ref{secKagomeFrames}.
	
	\subsection{Robustness of topological corner states in square frames}\label{secSquare_Robust}
	We analyze the robustness of higher-order topological corner states from three perspectives in this subsection. First we propose an analytical argument on the equivalence of the condition $\lvert C(\beta_tl_1) A(\beta_tl_2) \rvert = \lvert C(\beta_tl_2) A(\beta_tl_1) \lvert$, the emergence or disappearance of corner states, and degeneracy of candidate edge-state bands (i.e., closure of the bandgap of candidate edge states): these three events happen simultaneously. Second, we perform theoretical calculations: when defects are introduced into a square frame structure, the frequencies and localization lengths of topological corner states are computed using the dynamical matrix method. Finally, we conduct finite-element simulations to calculate the frequencies and localization lengths of topological corner states for the defect-inflicted structure. The results are then compared.
	
	First, we recall from Section~\refeq{secSquareCorner} that the frequencies of the corner states are given by Eq.~\eqref{eq5_1}, and the existence of the states are dictated by the relative magnitudes of $\left\lvert \frac{C(\beta_t l_1)}{A(\beta_t l_1)} \right\rvert$ and $\left\lvert \frac{C(\beta_t l_2)}{A(\beta_t l_2)} \right\rvert$ (Condition~\eqref{eq5_2}); thus, the condition for the emergence and disappearance of such corner states is exactly $\lvert C(\beta_tl_1) A(\beta_tl_2) \rvert = \lvert C(\beta_tl_2) A(\beta_tl_1) \lvert$. Continuously varying the geometric parameters $l_{1,2}$ does not affect the existence of corner states, as long as this condition is not triggered; moreover, the corner states do consistently reside within the bandgaps of candidate edge states, whose frequencies are given by Eq.~\eqref{eq5_10}. This is proved in Section~\ref{secSquareEdge}, where we demonstrate that when $\left\lvert \frac{C(\beta_tl_1)}{A(\beta_tl_1)}\right\rvert\neq\left\lvert\frac{C(\beta_tl_2)}{A(\beta_tl_2)}\right\lvert$, there exists one band of candidate edge states on each side of the corner-state frequency $\beta_t$, with a frequency range isolated from $\beta_t$. In contrast, when $\lvert C(\beta_tl_1) A(\beta_tl_2) \rvert = \lvert C(\beta_tl_2) A(\beta_tl_1) \lvert$, the boundaries of these candidate edge bands touch and become degenerate at $\beta_t$. Therefore, as long as the (candidate) edge bands remain nondegenerate, the existence of corner states are unchanged, indicating robustness.
	
	Analogous conclusions also hold for edge states: the emergence and disappearance of edge states are associated with the crossing of bulk bands, typically originating at point $X$ of the Brillouin zone. The band structure shown in Fig.~\ref{Figure_5} contains two such near-degenerate points, occurring at $\beta L = 9.45$ and $18.85$. From the dispersion curves of the square frame structure, it is evident that point $X$ of the Brillouin zone does not necessarily correspond to the band-edge frequencies of the bulk bands. Consequently, the frequency ranges of both corner states and edge states in higher-order topological frames may overlap with those of bulk states.
	
\begin{figure}[tb!] %6
		\centering
		\includegraphics[width=\linewidth]{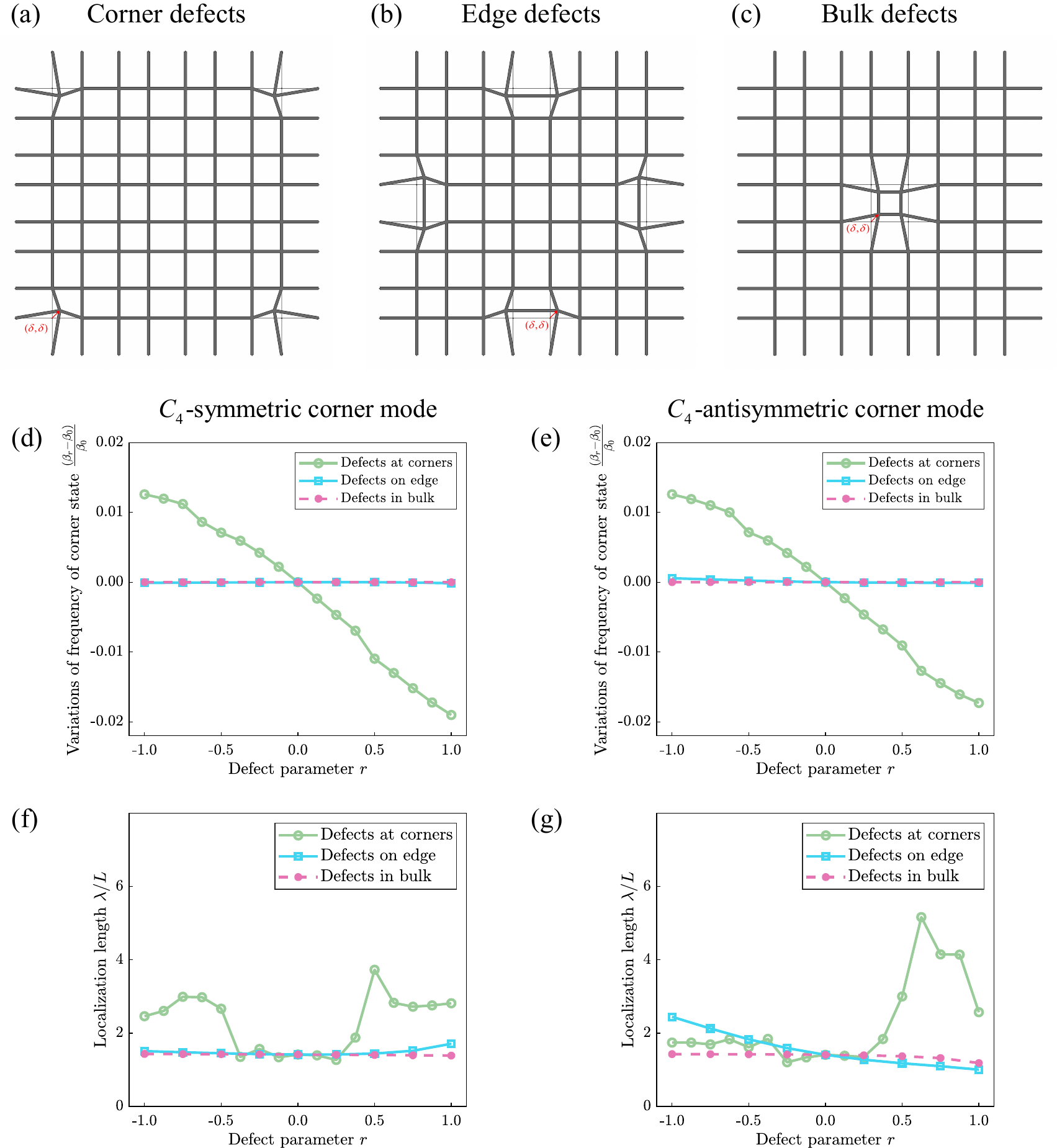}
		\caption{Theoretical calculations of topological corner states in square frames with defects. (a)--(c) Square frames with three types of defects, where the displacements of the joints at the defects are indicated by red arrows. (d)--(e) Frequency variations of the corner states with $C_4$-symmetry and $C_4$-antisymmetry, with respect to the magnitude of defect $\delta$ for different types of defects. Here $r$ is defined so that $\delta = r\cdot\lvert l_1-l_2 \rvert/4$. (f)--(g) Localization lengths of the $C_4$-symmetric and $C_4$-antisymmetric corner states.}
		\label{Figure_6}
	\end{figure}

	Next, we introduce geometric defects into the frame configuration, and calculate the frequency and localization characteristics of the higher-order corner states under such perturbations using the dynamical-matrix formulation. We consider three types of defects that preserve $C_{4v}$ symmetry: defects at the corners, on the edge, and in the bulk, as illustrated in Fig.~\ref{Figure_6}(a)--(c). In these configurations, the positions of specific joints in the frame are shifted, resulting in corresponding changes in the lengths of the adjacent beams. The displacements of joints in the horizontal and vertical directions are set as $\delta_x = \delta_y = r\cdot\lvert l_1-l_2 \rvert/4$, and varying the defect parameter $r$ changes the magnitude of the defect. For each perturbed configuration, the frequencies and localization lengths of the corner states are computed. Specifically, we analyze the two corner states that are either $C_4$-symmetric or $C_4$-antisymmetric, which correspond to Fig.~\ref{Figure_3}(a) and Fig.~\ref{Figure_3}(d) at $r=0$. To enhance the accuracy for calculating localization lengths, we construct a square frame consisting of $8\times 8$ unit cells and compute the vibration modes corresponding to the corner states, where we extract the rotational angles $\theta_\mathrm{A}^{(m,n)}$ ($(m,n)$ being the unit-cell index as shown in Fig.~\ref{Figure_1}) on sublattice A within the bottom-left quadrant of the structure, in order to avoid finite-size effects. A least-squares fit is performed on $\ln \bigl\lvert\theta_\mathrm{A}^{(m,n)} \bigr\rvert$ against the norm-1 distance to the corner, $(m+n-2)L$. The slope of the resulting fitted line is denoted as $k_s$, and the localization length is subsequently obtained as $\lambda=-1/k_s$.
	
	\begin{figure}[tb!] %7
		\centering
		\includegraphics[width=\linewidth]{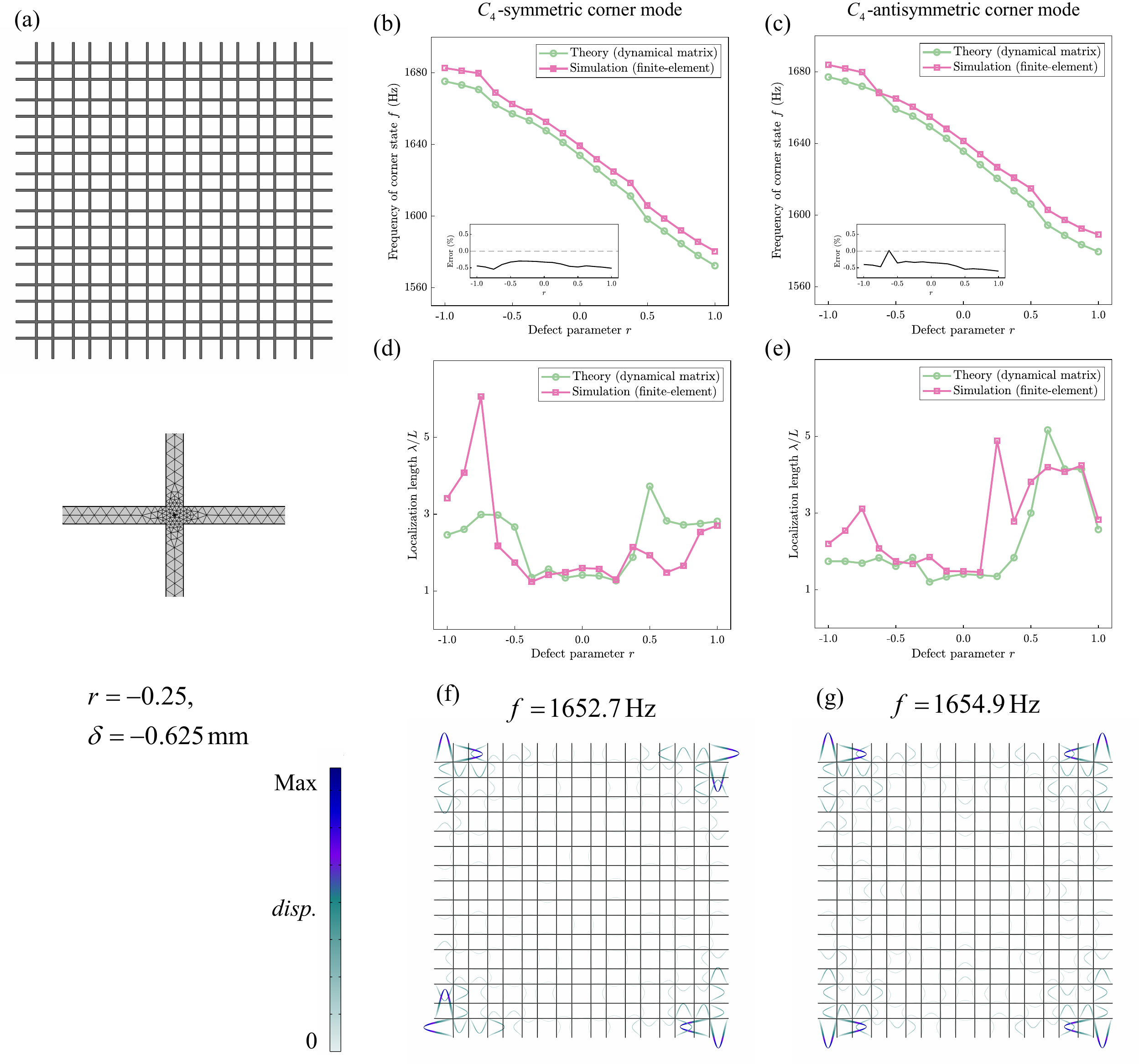}
		\caption{Finite-element analysis of topological corner states in a square frame with corner defects. (a) Upper: grid structure with $8\times 8$ cells used for simulation; the width of the beams is $1/40$ of $\min\{l_1, l_2\}$ (beams are thickened in the illustration for visual effects). Fixed boundary conditions are imposed on all outermost ends. Lower: typical mesh around a joint. (b)--(c) Frequencies of the $C_4$-symmetric and $C_4$-antisymmetric corner states from finite-element calculations as the magnitude of defects at the corners varies, together with a comparison to theoretical results. (d)--(e) Localization lengths of the $C_4$-symmetric and $C_4$-antisymmetric corner states from theoretical and finite-element results. (f)--(g) Corner states with a defect parameter of $r =-0.25$.}
		\label{Figure_7}
	\end{figure}
	
	Following the aforementioned approach, we analyze the three types of defects by varying the perturbation parameter within the range $r\in(-1,1)$. The frequencies of the symmetric corner state, calculated using the dynamical matrix method, are presented in Fig.~\ref{Figure_6}(d), while those of the antisymmetric corner state are shown in Fig.~\ref{Figure_6}(e). Additionally, the corresponding localization lengths for the symmetric and antisymmetric corner states are given in Fig.~\ref{Figure_6}(f)--(g). From Fig.~\ref{Figure_6}(d)--(e), it is seen that defects at the corners have the most significant effect on the frequencies of the corner states. Nevertheless, within the range $r\in(-0.4,0.4)$, the variation of frequencies does not exceed $2\%$. In contrast, the defects on the edge or in the bulk have a negligible influence on the frequencies of corner states. On the other hand, the localization lengths of the two corner states are qualitatively unchanged under perturbations with $r\in(-0.4,0.4)$, and remain small positive numbers even for $\lvert r\rvert$ up to $1$, as illustrated in Fig.~\ref{Figure_6}(f)--(g). It is concluded that the topological corner states are well localized and are robust against a wide range of $r$.
	
		Finally, we demonstrate finite-element simulations for robustness of the corner states of a square frame, for the purposes of investigating the influence of geometric defects on the frequencies and localization lengths of higher-order topological corner states. The simulations are conducted using the Solid Mechanics module of the software COMSOL Multiphysics. A set of free triangular mesh is constructed, and we ensure that each beam segment contains at least two layers of elements along the width direction (see the lower part of Fig.~\ref{Figure_7}(a)), thereby accurately capturing the bending deformations. The material is structural steel, with Young's modulus of $200\,\mathrm{GPa}$ and a density of $7850\,\mathrm{kg/m^3}$. %The finite-element method is employed to model the elastic frame structure with two primary objectives: first, to investigate the influence of geometric defects on the frequencies and localization lengths of higher-order topological corner states, thereby demonstrating the influence of such defects on the robustness of corner modes; and second, to validate the effectiveness of the theoretical framework proposed in this study for computing topological states in practical systems through elastodynamic simulations.

	For the configuration containing defects at the corners, the frequencies of the $C_4$-symmetric and $C_4$-antisymmetric topological modes obtained from finite-element simulations are presented in Fig.~\ref{Figure_7}(b)--(c) in relation to the defect parameter $r$. In these figures, a comparison is made between the finite-element and the theoretical results based on the dynamical matrix method described in this paper. The discrepancy between the two approaches is found to be within $1\%$ for the mode frequencies, thereby validating the accuracy of the proposed theoretical framework for obtaining the frequencies of topological states in such structures. The localization lengths of the symmetric and antisymmetric topological modes with respect to the defect parameter $r$, obtained from finite-element simulations, are shown in Fig.~\ref{Figure_7}(d)--(e). The topological corner states remain robust against a wide range of $r$. As an example, for $r=-0.25$, the vibration modes of the symmetric and antisymmetric corner states in the square grid-like frame are illustrated in Fig.~\ref{Figure_7}(f)--(g), demonstrating that the deflections of beams are predominantly localized at the corners.
	
	\subsection{Square frame heterostructure with corner states at the interface}\label{secSquare_Combine}
\begin{figure}[t!] %8
		\centering
		\includegraphics[width=\linewidth]{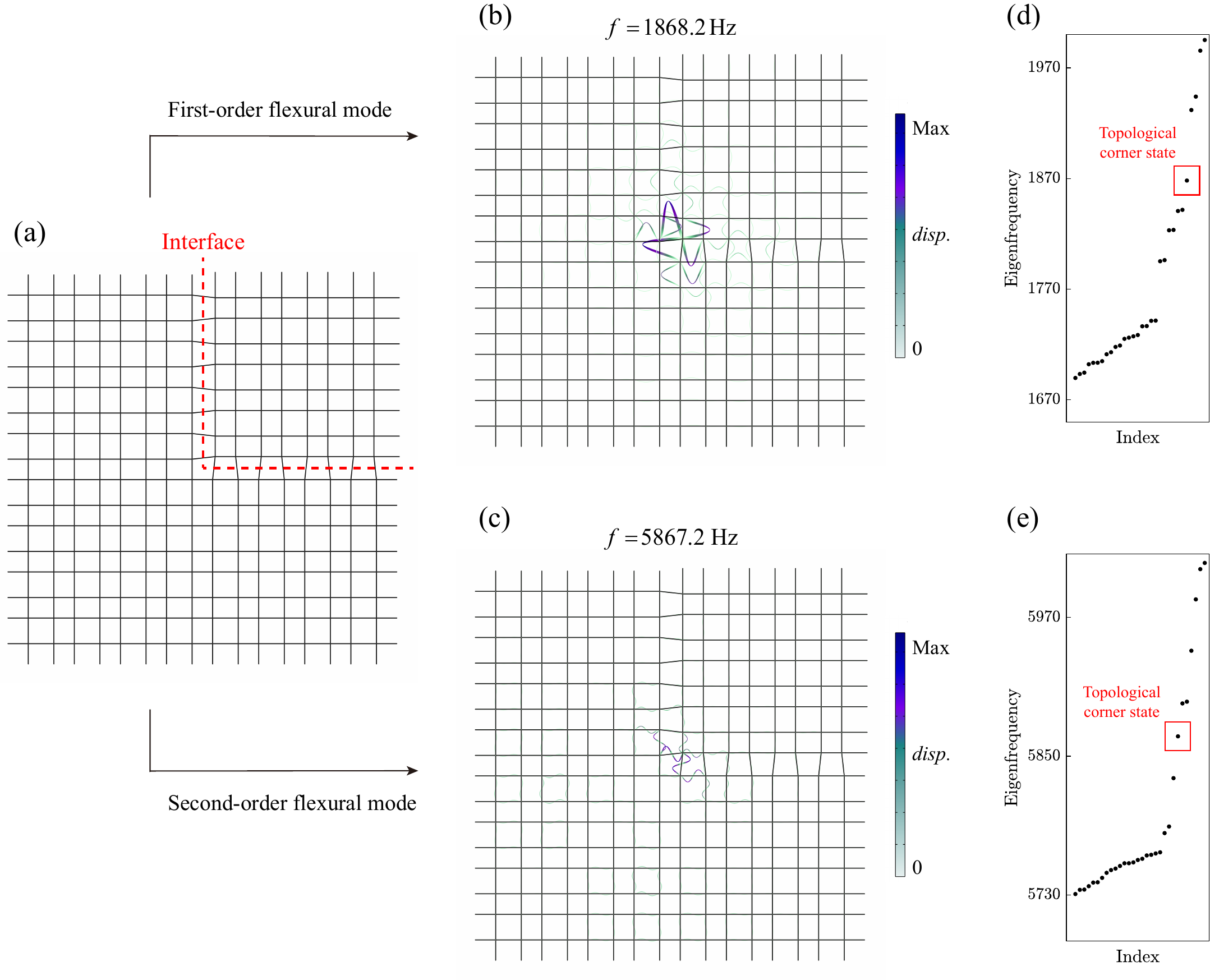}
		\caption{The topological corner states obtained at the interface in different frequency ranges. (a) A heterostructure obtained by connecting two lattice structures with different length parameters $(l_1,l_2)$. In simulation, the parameters are $(l_1,l_2) = (40\,\mathrm{mm}, 50\,\mathrm{mm})$ (upper-right part) and $(l_1,l_2) = (50\,\mathrm{mm}, 40\,\mathrm{mm})$ (remaining part). (b)--(c) Topological corner modes with the frequencies of $1868.2\,\mathrm{Hz}$ and $5867.2\,\mathrm{Hz}$ at the interface, where the beam segments are at their first and second flexural modes. (d)--(e) Mode spectrum around the frequency-isolated corner modes, with a difference in frequency of more than around $30\,\mathrm{Hz}$ from nearby modes.}
		\label{Figure_8}
	\end{figure}

	We consider combining two grid-like frame structures with distinct topological phases to form an interface as in Fig.~\ref{Figure_8}(a), which can host topological modes within common bandgaps.
	
\begin{figure}[t!] %9
		\centering
		\includegraphics[width=0.85\linewidth]{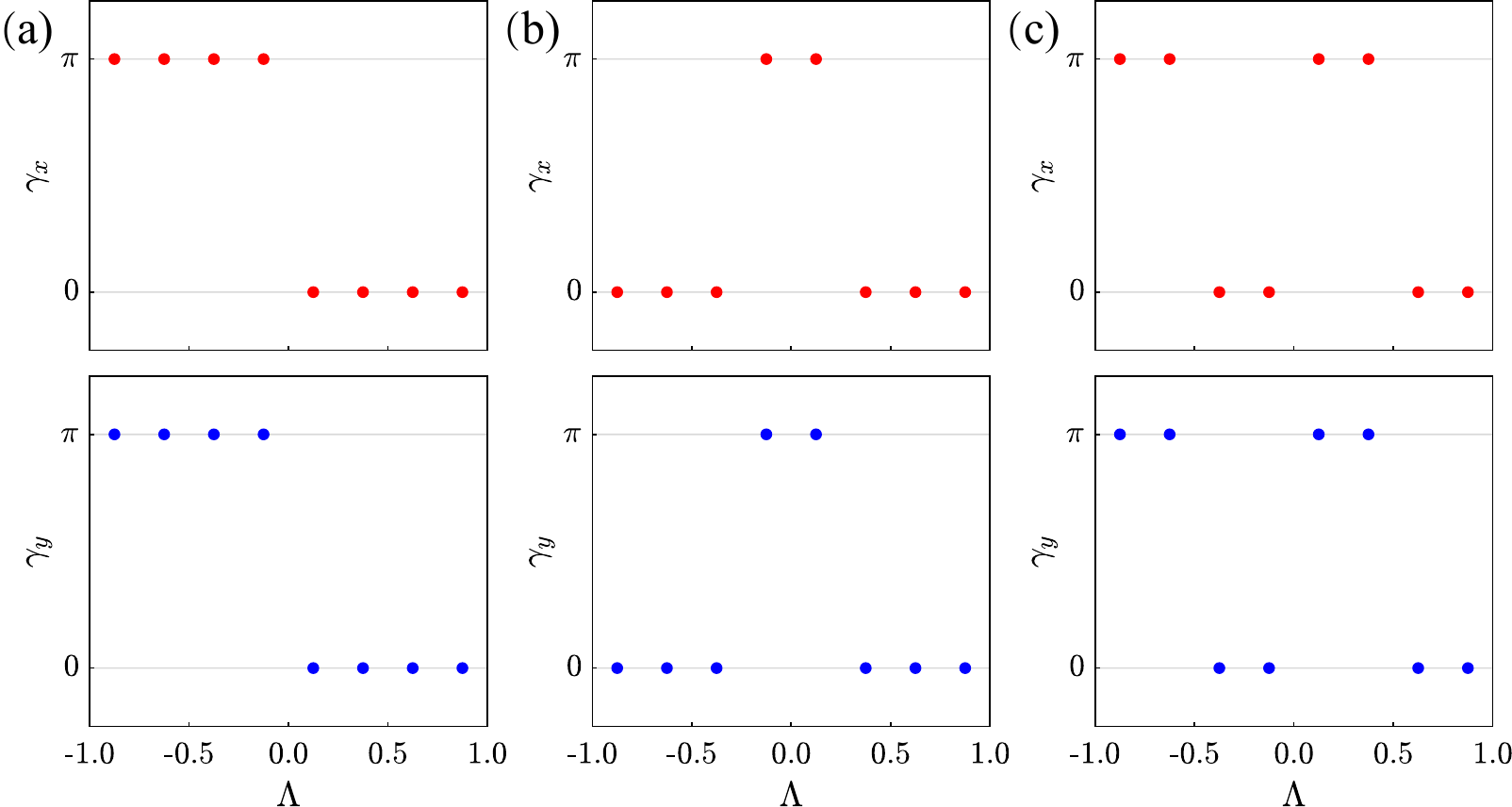}
		\caption{Numerical results of 2D multiband Zak phases $(\gamma_x,\gamma_y)$, with respect to the geometric parameter $\Lambda$, where $\Lambda$ is defined such that the lengths of beams are $ l_1 =(1+\Lambda) l_\mathrm{ave}$ and $ l_2 =(1-\Lambda) l_\mathrm{ave}$ (here $l_\mathrm{ave} = 45\,\mathrm{mm}$ is fixed). (a) $(\gamma_x^{(1)},\gamma_y^{(1)})$ for determining the existence of corner states at the interface near $\beta_t^{(1)}$. (b)--(c) $(\gamma_x^{(4)},\gamma_y^{(4)})$, $(\gamma_x^{(7)},\gamma_y^{(7)})$ for determining the existence of corner states near $\beta_t^{(2)}$ and $\beta_t^{(3)}$, respectively.
		}
		\label{Figure_9}
	\end{figure}

	Since exchanging the geometric parameters $(l_1, l_2) \leftrightarrow (l_2, l_1)$ simply amounts to a translation of the bulk (as long as no boundaries are concerned), the two frame structures obtained as such share identical bulk bands. As the continuum grid frames possess higher-order topological properties, topological corner states may emerge at the interface. The existence of topological corner states at the interface is determined by the difference between the bulk topological invariants of the two structures, in this case the 2D multiband Zak phases, defined as \citep{Zak1989,resta94}
	\begin{equation}
		(\gamma_x^{(n)}, \gamma_y^{(n)}) = \frac{L}{2\pi} \iint_{\mathrm{BZ}} \mathrm{d}k_x \,\mathrm{d}k_y \,\sum_{j=1}^n \langle w_j(\mathbf{k}) \rvert \mathrm{i} \nabla_\mathbf{k} \lvert w_j(\mathbf{k}) \rangle,
	\end{equation}
	where $\lvert w_j(\mathbf{k}) \rangle$ is the normalized in-cell displacement field of the $j$-th frequency band at wavevector $\mathbf{k}$; the sum is over the first $n$ bands; and the integral is over the first Brillouin zone. We calculate the 2D multiband Zak phases for square frame structures with different combinations of $(l_1, l_2)$, following the numerical scheme in \cite{resta94}; the results for $n=1$, $4$, $7$ which determine the existence of the corner states at the interface near $\beta_t^{(1)}$, $\beta_t^{(2)}$, $\beta_t^{(3)}$ are shown in Fig.~\ref{Figure_9}. It is seen that the first bulk band is topologically nontrivial with a topological phase of $(\pi,\pi)$ for $l_1<l_2$, and topologically trivial with a topological phase of $(0,0)$ for $l_1>l_2$, which leads to a topologically protected interface corner state above the first frequency band, upon combining two structures with interchanged parameters $l_{1,2}$. However, the topological phase becomes more complicated for higher-frequency bands, where multiple instances of phase transition occur, and exchanging $l_1$ and $l_2$ for the two bulks does not induce an interface corner state near $\beta_t^{(2m)}$ (corresponding to band number $n=6m-2$), as the topological phase $(\gamma_x^{(n)}, \gamma_y^{(n)})$ is unaltered in these cases. The numerical results are consistent with the analytical result that the topological phase transitions happen at $\left\lvert\frac{C(\beta_tl_1)}{A(\beta_tl_1)}\right\rvert=\left\lvert\frac{C(\beta_tl_2)}{A(\beta_tl_2)}\right\rvert$.
	
	We construct a heterostructure for simulation in the finite-element software COMSOL Multiphysics, as shown in Fig.~\ref{Figure_8}(a), which is formed by connecting a square frame with length parameters $(l_1,l_2) = (40\,\mathrm{mm}, 50\,\mathrm{mm})$ in the upper-right part and another square frame with $(l_1,l_2) = (50\,\mathrm{mm}, 40\,\mathrm{mm})$ occupying the remaining part. The finite-element analysis reveals an isolated topological corner state at $1868.2\,\mathrm{Hz}$ as shown in Fig.~\ref{Figure_8}(b), where the beam segments predominantly exhibit first-order flexural mode shapes. Its frequency lies between the first and fourth bulk bands, and is slightly above $\beta_t^{(1)}$. Going to higher-frequency regions, no such corner state is found around $\beta_t^{(2)}$, and the next occurrence of a corner state lies around $\beta_t^{(3)}$, above the seventh band; the beam segments vibrate in second-order flexural mode shapes (Fig.~\ref{Figure_8}(c)). Such behavior is consistent with numerical results of the topological phases in Fig.~\ref{Figure_9}, manifesting the bulk--boundary correspondence for higher-order topological heterostructures. The structure possesses more corner states in even higher-frequency regions. These corner states are separated from the nearby modes as shown in Fig.~\ref{Figure_8}(d)--(e), and as a consequence, such heterostructures have potential applications for robust waveguiding \citep{PhysRevLett.122.233903,zhang_low-threshold_2020}.
	
	\section{Topological kagome grid-like frames}\label{secKagomeFrames}
	In this section, we consider the kagome frame structure and analyze its (first-order and higher-order) topological properties. The kagome frame structure consists of beam segments with alternating lengths $l_1$ and $l_2$ along each of the three non-orthogonal directions; as shown in Fig.~\ref{Figure_10}(a), the structure contains two different types of equilateral triangles formed by beam segments, one pointing upwards with side length $l_1$, and the other pointing downwards with side length $l_2$. The unit cell is taken to include the upward-pointing triangle, whose three rigid joints lie at sublattices A, B and C, respectively. Therefore, the lengths of intracell beam segments are $l_1$, and the lengths of intercell beam segments $l_2$. The whole structure contains $N(N+1)/2$ unit cells in total, where $N$ is the number of unit cells along each side. All the exterior ends of the outermost beam segments are clamped ends. Each unit cell is labeled by $(m, n)$, so that the position vector of the cell is $\mathbf{r} = mL \mathbf{e}_1 + nL \mathbf{e}_2$ (where $L=l_1+l_2$ is the lattice constant, and $\mathbf{e}_1$, $\mathbf{e}_2$ are unit vectors along lattice directions, depicted in Fig.~\ref{Figure_10}(a)). We investigate the existence of edge states and higher-order topological corner states of the kagome frame, and determine the eigenfrequencies of the corner states, edge states, and bulk states in the frequency spectrum.
	
	We first take a unit cell of the kagome frame and perform a Bloch-wave analysis, through which the dynamical equation is obtained as
	\begin{equation}	\label{eq5_15add}
		H^{\text{kagome}}_{\text{Bloch}} \lvert \theta \rangle=0,
	\end{equation}
	where $\lvert\theta\rangle = (\theta_\mathrm{A}, \theta_\mathrm{B}, \theta_\mathrm{C})^T$ denotes the rotation angles at the three rigid joints in one unit cell, and the Bloch dynamical matrix is
	\begin{equation}
		\label{eq5_16_add2}
		\small
		H^{\text{kagome}}_{\text{Bloch}} =
		\begin{bmatrix}
			\displaystyle 2\sum\limits_{i=1}^2 \frac{B(\beta l_i)}{A(\beta l_i)} & \displaystyle -\frac{C(\beta l_1)}{A(\beta l_1)} - \frac{C(\beta l_2)}{A(\beta l_2)} \mathrm{e}^{-\mathrm{i}k_1L} & \displaystyle -\frac{C(\beta l_1)}{A(\beta l_1)} - \frac{C(\beta l_2)}{A(\beta l_2)} \mathrm{e}^{-\mathrm{i}k_2L}
			\\[9pt]
			\displaystyle -\frac{C(\beta l_1)}{A(\beta l_1)} - \frac{C(\beta l_2)}{A(\beta l_2)} \mathrm{e}^{\mathrm{i}k_1L} & \displaystyle 2\sum\limits_{i=1}^2 \frac{B(\beta l_i)}{A(\beta l_i)} & \displaystyle -\frac{C(\beta l_1)}{A(\beta l_1)} - \frac{C(\beta l_2)}{A(\beta l_2)} \mathrm{e}^{\mathrm{i}(k_1-k_2)L}
			\\[9pt]
			\displaystyle -\frac{C(\beta l_1)}{A(\beta l_1)} - \frac{C(\beta l_2)}{A(\beta l_2)} \mathrm{e}^{\mathrm{i}k_2 L} & \displaystyle -\frac{C(\beta l_1)}{A(\beta l_1)} - \frac{C(\beta l_2)}{A(\beta l_2)} \mathrm{e}^{\mathrm{i}(k_2-k_1)L} & \displaystyle 2\sum\limits_{i=1}^2 \frac{B(\beta l_i)}{A(\beta l_i)}
		\end{bmatrix}.
	\end{equation}
	Here $k_1 = \mathbf{k} \cdot \mathbf{e}_1 = k_x$, and $k_2 = \mathbf{k} \cdot \mathbf{e}_2 = (1/2)k_x + (\sqrt{3}/2)k_y$. $k_1L$ denotes the phase difference between the joint rotational angles of the two neighboring unit cells on the same sublattice along the direction $\mathbf{e}_1$, and $k_2L$ denotes the phase difference along $\mathbf{e}_2$.
	
	Similarly, for a finite-sized kagome grid-like frame, the dynamical matrix is denoted as $H^{\text{kagome}}$, whose main-diagonal elements are $2\sum\limits_{i=1}^2 \frac{B(\beta l_i)}{A(\beta l_i)}$, and off-diagonal elements are $-\frac{C(\beta l_1)}{A(\beta l_1)}$ or $-\frac{C(\beta l_2)}{A(\beta l_2)}$ depending on the length of the beam segment corresponding to the entry position.
	
	\subsection{Higher-order topological corner states in kagome frames}
	\begin{figure}[tb!] %10
		\centering
		\includegraphics[width=\linewidth]{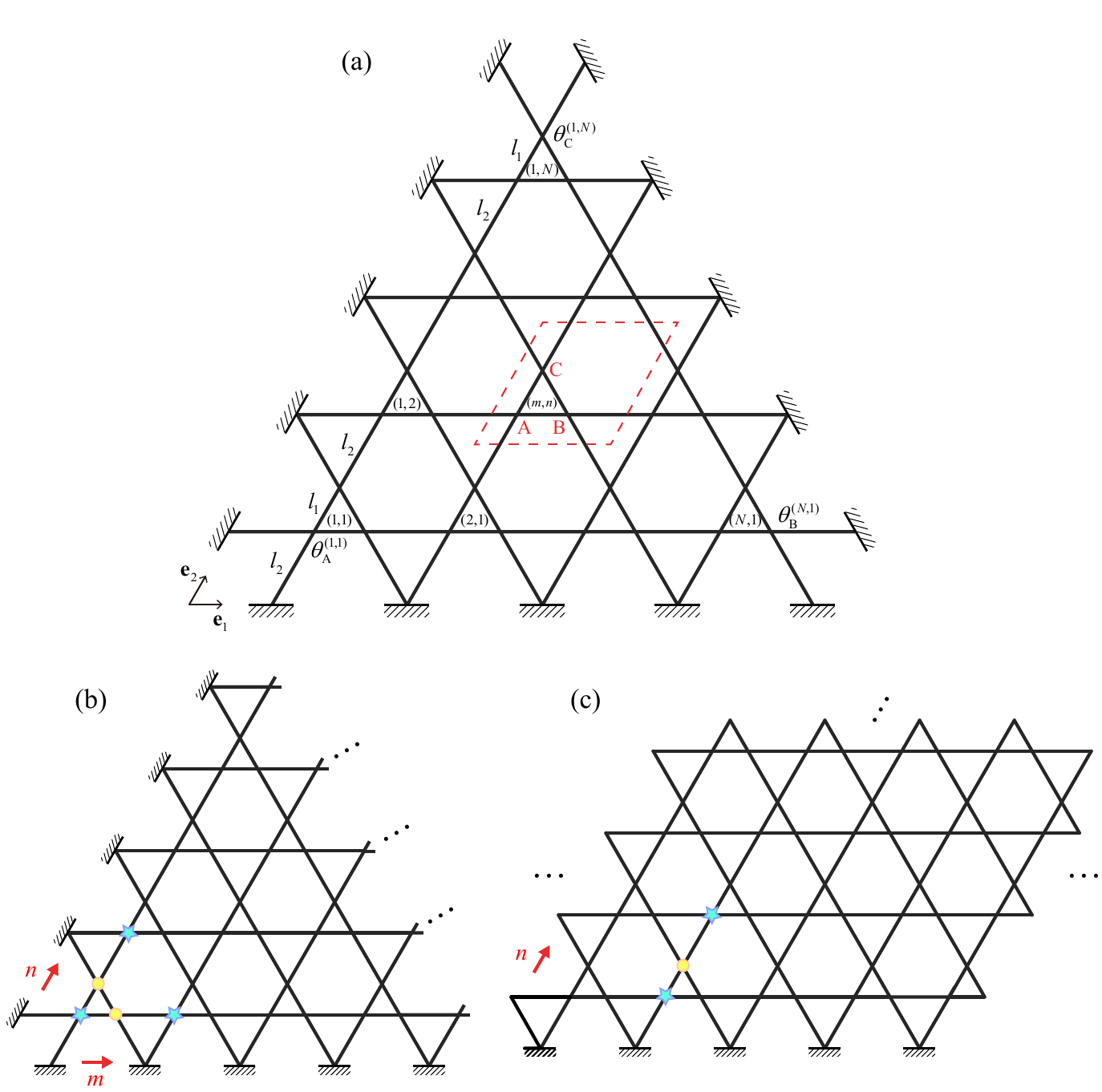}
		\caption{(a) Kagome grid-like frame without translational displacements at joints, where a unit cell is marked by red dashed lines. (b) Schematic illustration of the solution process of the existence condition of corner states, i.e., Eqs.~\eqref{eq5_16}--\eqref{eq5_17}. (c) Schematic illustration of the solution process of the existence condition of edge states, i.e., Eqs.~\eqref{eq5_24}--\eqref{eq5_25add}.}
		\label{Figure_10}
	\end{figure}
\begin{figure}[tbp!] %11
		\centering
		\includegraphics[width=0.95\linewidth]{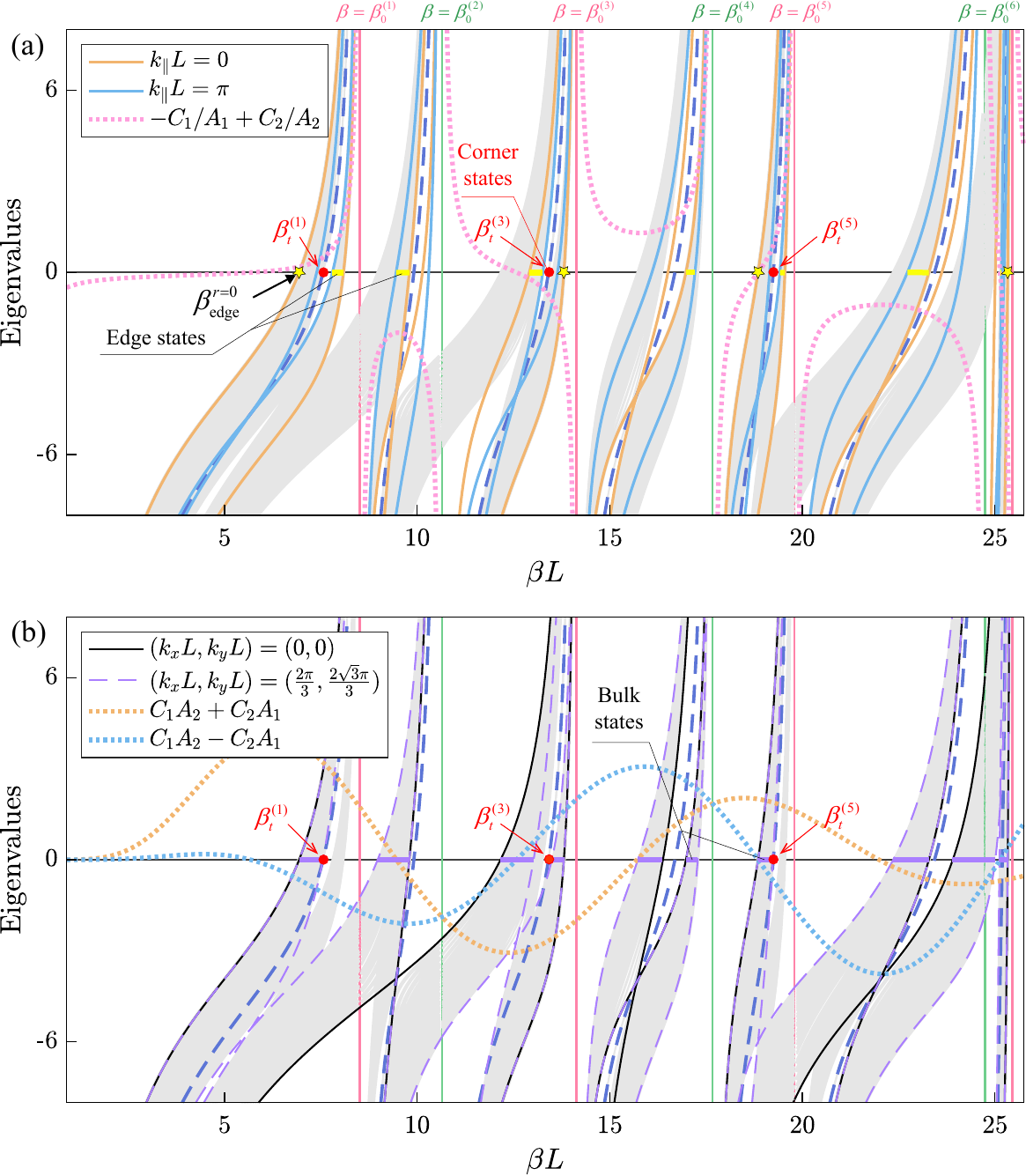}
		\caption{Eigenvalue spectrum of a topological kagome grid-like frame, with $l_1=40\,\mathrm{mm}$, $l_2=50\,\mathrm{mm}$. Gray curves represent eigenvalues of the dynamical matrix $H^{\mathrm{kagome}}$ of the finite-sized frame, and all the intersections of gray curves and the black horizontal line $\lambda=0$ constitute the frequency spectrum of the frame. Pink (green) vertical lines represent the roots of equation $A(\beta_0 l_{2(1)})=0$. The intersections of the dark blue dashed curve with $\lambda=0$ marked by red arrows indicate the frequencies of topological corner states. (a) Edge states are indicated by yellow line segments, where the values of the functions $2 [B(\beta l_1)/A(\beta l_1)+B(\beta l_2)/A(\beta l_2)]$ (dark blue dashed curves) and $-C(\beta l_1)/A(\beta l_1)+C(\beta l_2)/A(\beta l_2)$ (pink dotted curves) are of the same sign. Edge states also exist at frequencies indicated by yellow stars. (b) Bulk bands correspond to regions on the $\lambda=0$ line bounded by adjacent black solid curves and purple dashed curves.}
		\label{Figure_11}
	\end{figure}
\begin{figure}[tb!] %12
		\centering
		\includegraphics[width=\linewidth]{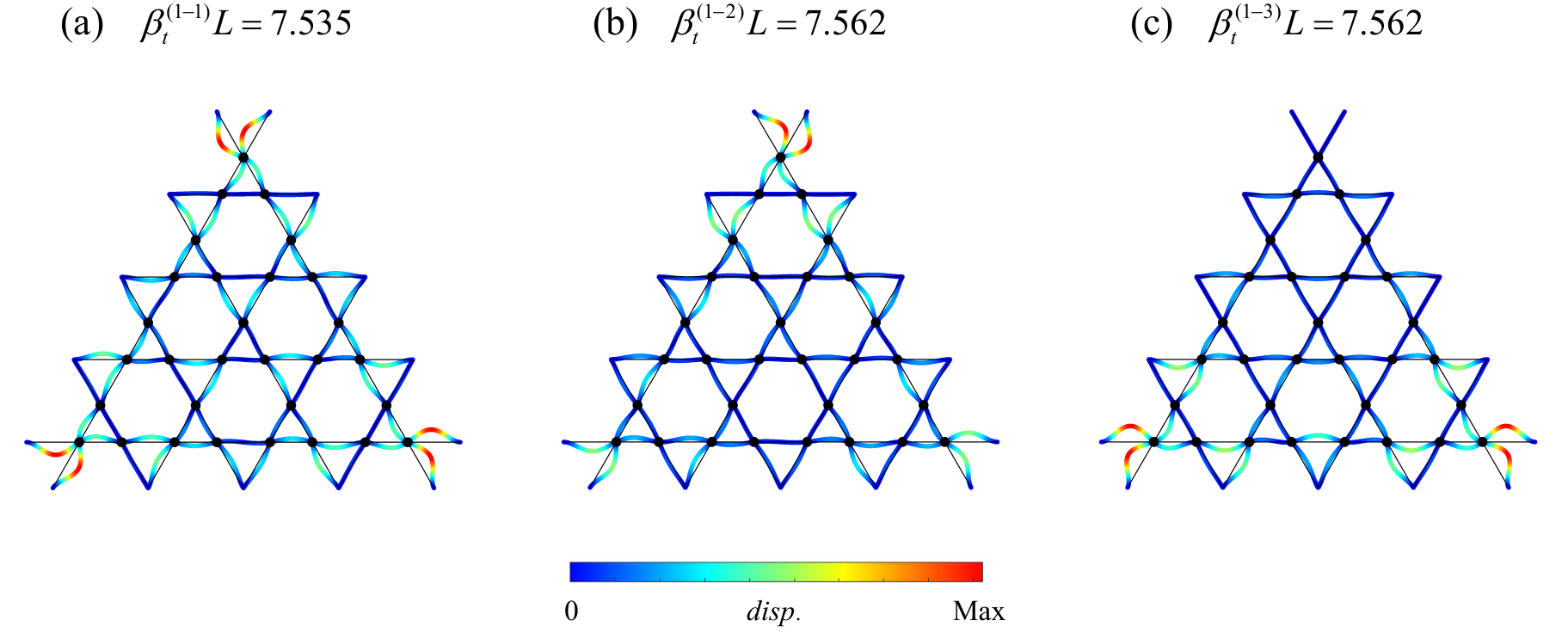}
		\caption{Topological corner modes in kagome grid-like frames. The modes are triply degenerate at frequency $\beta_{t}^{(1)}$, and localized at the corners.}
		\label{Figure_12}
	\end{figure}

	Since the kagome lattice has generalized chiral symmetry \citep{kagome,10.1002/adma.202312421} (after the main-diagonal elements of the matrix $H_{\mathrm{Bloch}}^{\mathrm{kagome}}$ become identically zero by subtracting a scalar matrix), the frequencies of topological corner states in a finite kagome grid-like frame with fixed-end boundaries must satisfy \citep{SUN2025105935}
	\begin{equation}
		\label{eq5_16}
		2 \left[\frac{B(\beta_t l_1)}{A(\beta_t l_1)}+\frac{B(\beta_t l_2)}{A(\beta_t l_2)}\right] = 0.
	\end{equation}
	Meanwhile, if and only if \citep{SUN2025105935}
	\begin{equation}
		\label{eq5_17}
		\left \lvert\frac{C(\beta_t l_1)}{A(\beta_t l_1)}\right \rvert<\left \lvert\frac{C(\beta_t l_2)}{A(\beta_t l_2)}\right \rvert,
	\end{equation}
	topological corner states exist at $\beta_t$. The eigenfrequency spectrum of an example kagome frame with $l_1=40\,\mathrm{mm}$ and $l_2=50\,\mathrm{mm}$ is depicted in Fig.~\ref{Figure_11}. The frequencies of topological corner states are marked with red arrows in Fig.~\ref{Figure_11}. As shown in Fig.~\ref{Figure_11}(b), the ordinates of the blue and orange dotted curves have opposite signs at frequencies $\beta_t^{(1)}$, $\beta_t^{(3)}$ and $\beta_t^{(5)}$, ensuring the existence of topological corner states at these frequencies; this method of judgment is exactly equivalent to the criterion~\eqref{eq5_17}. Although certain frequencies of topological corner states may lie within bulk bands, for example $\beta_t^{(1)}$, the topological corner states still exist. The corner modes obtained theoretically are shown in Fig.~\ref{Figure_12}.
	
	
	In the following we illustrate the mode characteristics of the above topological corner states in Fig.~\ref{Figure_12}, including localization, using an analytical approach. %prove that topological corner states exist and are localized at corners if and only if the conditions~\eqref{eq5_16} and \eqref{eq5_17} both hold. 
	Consider a semi-infinite kagome lattice with one corner boundary as shown in Fig.~\ref{Figure_10}(b). Inspired by Bloch's theorem, we let the corner-state solution be of the form
	\begin{equation}
		\label{eq5_18}
		\theta_\mathrm{A}^{(m,n)} = \Theta_\mathrm{A} \mathrm{e}^{\mathrm{i}\mathbf{k}\cdot\mathbf{r}} = \Theta_\mathrm{A} \mathrm{e}^{\mathrm{i}(mk_1L+nk_2L)};
	\end{equation}
	similarly, let
	\begin{equation}
		\label{eq5_18_add}
		\theta_\mathrm{B}^{(m,n)}=\Theta_\mathrm{B} \mathrm{e}^{\mathrm{i}(mk_1L+nk_2L)}, \quad \theta_\mathrm{C}^{(m,n)}=\Theta_\mathrm{C} \mathrm{e}^{\mathrm{i}(mk_1L+nk_2L)}.
	\end{equation}
	
	Because we wish to solve the corner state which is localized at the corner and exponentially decays along both directions $\mathbf{e}_1$ and $\mathbf{e}_2$, we define the decay  coefficient as $\mathrm{e}^{\mathrm{i}k_1L}=r_1$ and $\mathrm{e}^{\mathrm{i}k_2L}=r_2$ (here $k_1$, $k_2$ are no longer real numbers and have a nonzero imaginary part), and hence
	\begin{equation}
		\label{eq5_19}
		\theta_\mathrm{A(B,C)}=\Theta_\mathrm{A(B,C)} r_1^m r_2^n.
	\end{equation}
	We compare the balance equations of the bending moments at joint A in unit cells $(1,1)$ and $(2,1)$ (labeled by blue stars in Fig.~\ref{Figure_10}(b)):
	\begin{align}
		0 &= 2 \left[ \frac{B(\beta l_1)}{A(\beta l_1)} + \frac{B(\beta l_2)}{A(\beta l_2)} \right] \theta_\mathrm{A}^{(1,1)} - \frac{C(\beta l_1)}{A(\beta l_1)} \theta_\mathrm{B}^{(1,1)} - \frac{C(\beta l_1)}{A(\beta l_1)} \theta_\mathrm{C}^{(1,1)}, \label{eq:kagomeSiteA11} \\
		0 &= 2 \left[ \frac{B(\beta l_1)}{A(\beta l_1)} + \frac{B(\beta l_2)}{A(\beta l_2)} \right] \theta_\mathrm{A}^{(2,1)} - \frac{C(\beta l_1)}{A(\beta l_1)} \theta_\mathrm{B}^{(2,1)} - \frac{C(\beta l_1)}{A(\beta l_1)} \theta_\mathrm{C}^{(2,1)} - \frac{C(\beta l_2)}{A(\beta l_2)} \theta_\mathrm{B}^{(1,1)}. \label{eq:kagomeSiteA21}
	\end{align}
	Considering Eq.~\eqref{eq5_19}, we obtain $\theta^{(1,1)}_\mathrm{B}=0$. Thus $\Theta_\mathrm{B}=0$, and it follows that
	\begin{equation}
		\label{eq5_20}
		\theta_{\mathrm{B}}^{(m,n)}=0.
	\end{equation}
	In a similar way, we compare the balance equations of the bending moments at joint A in unit cells $(1,1)$ and $(1,2)$ (labeled by blue stars), and $\theta^{(1,1)}_\mathrm{C}=0$ is obtained. Thus $\Theta_\mathrm{C}=0$, and it follows that
	\begin{equation}
		\label{eq5_21}
		\theta_{\mathrm{C}}^{(m,n)}=0.
	\end{equation}
	
	Then, for joint C in unit cell $(m,n)$ (marked by one of the yellow stars), the balance equation of the bending moments is
	\begin{equation}
		\label{eq5_22}
		t_2\theta_\mathrm{A}^{(m,n+1)}+t_1\theta_\mathrm{A}^{(m,n)}=0,
	\end{equation}
	where $t_1$ is defined as $-C(\beta l_1)/A(\beta l_1)$, and $t_2$ is defined as $-C(\beta l_2)/A(\beta l_2)$ (note that we have used Eqs.~\eqref{eq5_20} and \eqref{eq5_21}); here $\beta=\beta_t$ that satisfies Eq.~\eqref{eq5_16}. For joint B in unit cell $(m,n)$, the balance equation of the bending moments is
	\begin{equation}
		\label{eq5_23}
		t_2\theta_\mathrm{A}^{(m+1,n)}+t_1\theta_\mathrm{A}^{(m,n)}=0.
	\end{equation}
	By substituting Eq.~\eqref{eq5_19} into Eqs.~\eqref{eq5_22} and \eqref{eq5_23}, we obtain
	\[r_2=r_1=-\frac{t_1}{t_2}.
	\]
	Thus,
	\begin{equation}
		\label{eq5_23add}
		{\theta_\mathrm{A(B,C)}^{(m,n)}}=\Theta_\mathrm{A(B,C)} \left(-\frac{t_1}{t_2}\right)^{m+n}.
	\end{equation}
	When $\lvert r_{1,2}\rvert<1$, i.e., $\lvert t_1\rvert<\lvert t_2\rvert$, such a state is localized at the corner, and decays exponentially away from the corner. In this case, the solution $\lvert \theta \rangle$ given by Eqs.~\eqref{eq5_20}, \eqref{eq5_21} and \eqref{eq5_23add} satisfies the balance equations of the bending moments at all joints in the semi-infinite structure, thus indeed a feasible solution for the corner state.
	
	\subsection{Edge states in kagome frames}
	In this subsection, we first present the frequency ranges of edge states in kagome grid-like frames, and then give the proofs of the theoretical results.
	
\begin{figure}[tb!] %13
		\centering
		\includegraphics[width=\linewidth]{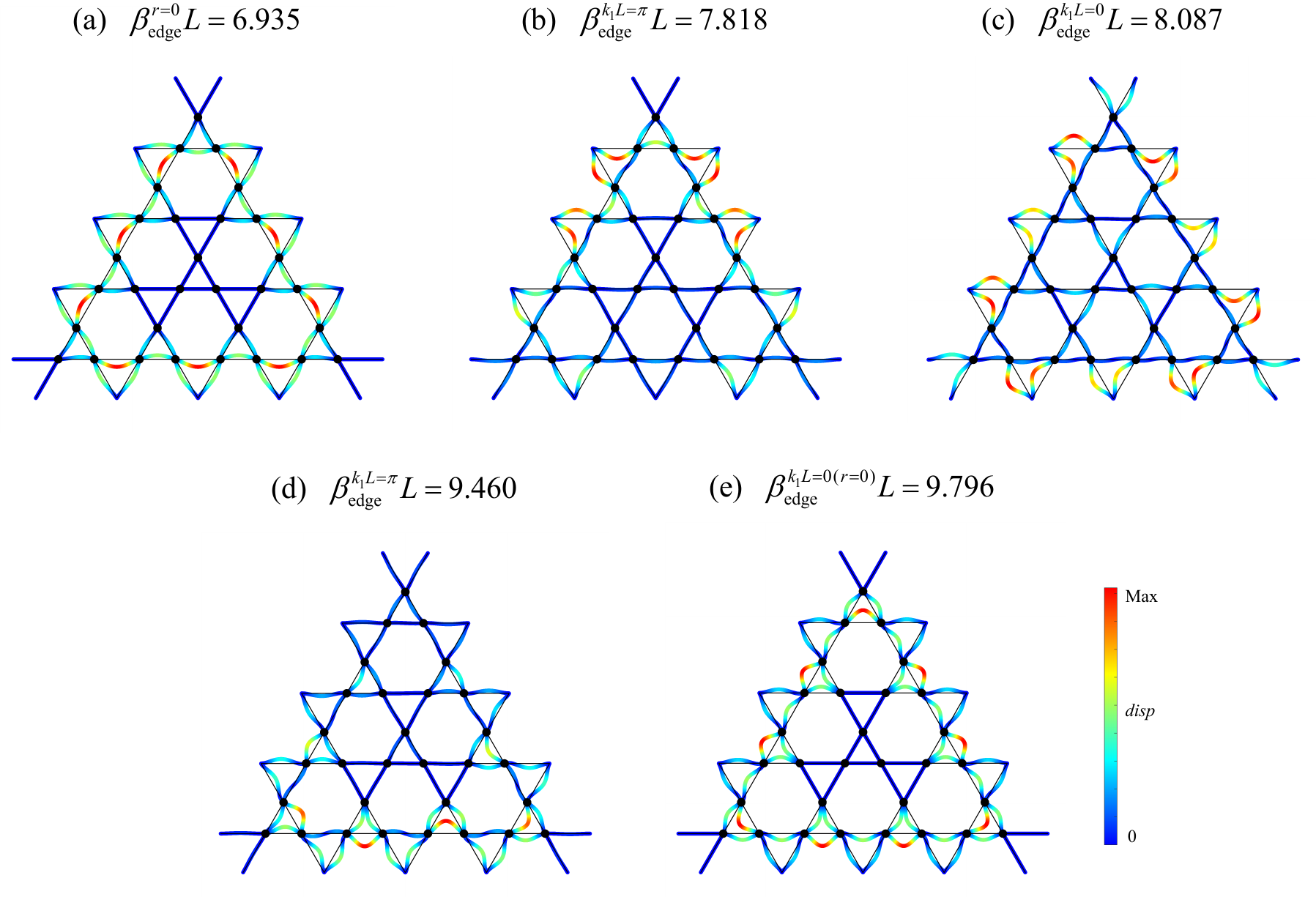}
		\caption{Edge modes of a kagome grid-like frame. (a) Edge state at frequency $\beta^{r=0}_\mathrm{edge}$. (b)--(c) Edge states corresponding to $k_{\parallel}L=\pi$ and $k_{\parallel}L=0$ in the interval $(\beta_t^{(1)},\beta_0^{(1)})$ shown in Fig.~\ref{Figure_11}. (d)--(e) Edge states corresponding to $k_{\parallel}L=\pi$ and $k_{\parallel}L=0$ in the interval $(\beta_0^{(1)},\beta_t^{(2)})$ shown in Fig.~\ref{Figure_11}.}
		\label{Figure_13}
	\end{figure}

	The frequencies of edge states in kagome frames satisfy
	\begin{equation}
		\label{eq5_24}
		2\left[\frac{B(\beta l_1)}{A(\beta l_1)}+\frac{B(\beta l_2)}{A(\beta l_2)}\right]=\pm \left\lvert\frac{C(\beta l_1)}{A(\beta l_1)} + \frac{C(\beta l_2)}{A(\beta l_2)} \exp(\mathrm{i}k_{\parallel}L)\right\rvert;
	\end{equation}
	Meanwhile, if and only if
	\begin{equation}
		\label{eq5_25}
		2\left[\frac{B(\beta l_1)}{A(\beta l_1)}+\frac{B(\beta l_2)}{A(\beta l_2)}\right]\cdot \left[-\frac{C(\beta l_1)}{A(\beta l_1)}+\frac{C(\beta l_2)}{A(\beta l_2)}\right]>0
	\end{equation}
	or
	\begin{equation}
		\label{eq5_25add}
		2\left[\frac{B(\beta l_1)}{A(\beta l_1)}+\frac{B(\beta l_2)}{A(\beta l_2)}\right]= -\frac{C(\beta l_1)}{A(\beta l_1)}-\frac{C(\beta l_2)}{A(\beta l_2)}
	\end{equation}
	do edge states exist at the frequencies $\beta$ satisfying Eq.~\eqref{eq5_24}. In Fig.~\ref{Figure_11}(a), the range of $\beta$ corresponding to Eq.~\eqref{eq5_24} appears as the set of points on $\lambda=0$ between a pair of neighboring blue and orange solid curves, in the intervals $(\beta_0^{(n-1)},\beta_t^{(n)})$ and $(\beta_t^{(n)},\beta_0^{(n)})$. This set of frequencies (i.e., solutions to Eq.~\eqref{eq5_24}) have the same range with the corresponding set of frequencies for the square frame as given in Eq.~\eqref{eq5_10}, and hence the same existence property (i.e., existence within each interval $(\beta_0^{(n-1)},\beta_t^{(n)})$ and $(\beta_t^{(n)},\beta_0^{(n)})$) follows. Although the candidate frequencies of edge states are the same, the existence condition of edge states for the kagome frame (the expressions~\eqref{eq5_25} and \eqref{eq5_25add}) is quite different from that for the square frame (the expression~\eqref{eq5_10a}). First, we examine the set of frequencies that satisfy the conditions~\eqref{eq5_24} and \eqref{eq5_25}. In Fig.~\ref{Figure_11}(a), the pink dotted curves represent function $\left[-\frac{C(\beta l_1)}{A(\beta l_1)}+\frac{C(\beta l_2)}{A(\beta l_2)}\right]$, and dark blue dashed curves represent function $2 \left[\frac{B(\beta l_1)}{A(\beta l_1)}+\frac{B(\beta l_2)}{A(\beta l_2)}\right]$. For each point set between a pair of blue and orange solid curves, when the values of the two functions $\left[-\frac{C(\beta l_1)}{A(\beta l_1)}+\frac{C(\beta l_2)}{A(\beta l_2)}\right]$ and $2 \left[\frac{B(\beta l_1)}{A(\beta l_1)}+\frac{B(\beta l_2)}{A(\beta l_2)}\right]$ have the same sign, the condition~\eqref{eq5_25} holds, and thus edge states exist at these $\beta$. Therefore, edge states exist at the frequency intervals marked by yellow line segments in Fig.~\ref{Figure_11}(a). Then, the solutions $\beta^{r=0}_\mathrm{edge}$ that satisfy Eq.~\eqref{eq5_25add} but do not lie within the intervals of the yellow line segments are indicated by yellow stars. Several examples are taken for illustration: selected edge states at the frequency $\beta^{r=0}_\mathrm{edge}$, in the interval $(\beta_t^{(1)},\beta_0^{(1)})$, and in the interval $(\beta_0^{(1)},\beta_t^{(2)})$ (see Fig.~\ref{Figure_11}(a) for details) are demonstrated in Fig.~\ref{Figure_13}(a), (b)--(c), and (d)--(e), respectively. The bending deformations are localized near the edges of the structure, which is typical for edge states.
	
	In the following, we prove the conditions \eqref{eq5_24} and \eqref{eq5_25} for the existence of edge states. Consider a semi-infinite kagome lattice as shown in Fig.~\ref{Figure_10}(c), where we employ Eqs.~\eqref{eq5_18} and \eqref{eq5_18_add} and attempt to solve edge states whose energy are localized on the boundary and decay exponentially into the bulk. Letting $\mathrm{e}^{\mathrm{i}k_2L}=r$, we obtain
	\begin{equation}
		\label{eq5_26}
		\theta_\mathrm{A(B,C)}=\Theta_\mathrm{A(B,C)} \mathrm{e}^{\mathrm{i}mk_1L} r^n.
	\end{equation}
	For edge states, the component $k_1$ along the edge is real. We compare the the balance equations of the bending moments at joint A in unit cells $(m,1)$ and $(m,2)$ (marked by blue stars in Fig.~\ref{Figure_10}(c)), which leads to $\theta^{(m,1)}_\mathrm{C}=0$. Thus $\Theta_\mathrm{C}=0$, and it follows that
	\begin{equation}
		\label{eq5_27}
		\theta_{\mathrm{C}}^{(m,n)}=0.
	\end{equation}
	Substituting Eq.~\eqref{eq5_27} into the balance equations of the bending moments for joints A and B in unit cell $(m,n)$, we obtain
	\begin{align}
		\label{eq5_28}
		\epsilon\theta_{\mathrm{A}}^{(m,n)} &= t_1\cdot \theta_{\mathrm{B}}^{(m,n)}+t_2\cdot \theta_{\mathrm{B}}^{(m,n)}\mathrm{e}^{-\mathrm{i}k_1L},\\
		\label{eq5_29}
		\epsilon\theta_{\mathrm{B}}^{(m,n)} &= t_1\cdot \theta_{\mathrm{A}}^{(m,n)}+t_2\cdot \theta_{\mathrm{A}}^{(m,n)}\mathrm{e}^{\mathrm{i}k_1L},
	\end{align}
	where $\epsilon\equiv-2\left[\frac{B(\beta l_1)}{A(\beta l_1)}+\frac{B(\beta l_2)}{A(\beta l_2)}\right]$, which is also the eigenvalue of the dynamical matrix with all diagonal elements set to zero; $t_1\equiv-C(\beta l_1)/A(\beta l_1)$, and $t_2\equiv-C(\beta l_2)/A(\beta l_2)$. By combining Eqs.~\eqref{eq5_28} and \eqref{eq5_29}, we obtain the candidate frequencies where edge states of the kagome frame may occur, that is, the candidate frequencies are solutions to
	\begin{equation}
		\label{eq5_29add}
		\lambda_\mathrm{edge}^\mathrm{kagome}=-\epsilon \pm \lvert t_1+t_2\mathrm{e}^{\mathrm{i}k_1L}\rvert=0,
	\end{equation}
	which are precisely the solutions to Eq.~\eqref{eq5_24}; moreover, we obtain
	\begin{equation}
		\label{eq5_30add}
		\theta_{\mathrm{B}}^{(m,n)}=\theta_{\mathrm{A}}^{(m,n)}\mathrm{e}^{\mathrm{i}\phi},
	\end{equation}
	where
	\begin{equation}
		\label{eqphiDef}
		\phi=\arg{\frac{t_1+t_2\mathrm{e}^{\mathrm{i}k_1L}}{\epsilon}}.
	\end{equation}
	
	Next, for joint C in unit cell $(m,n)$ (highlighted with a yellow star in Fig.~\ref{Figure_10}(c)), the balance equation of the bending moments is
	\begin{equation}
		\label{eq5_30}
		t_2\left[\theta_{\mathrm{A}}^{(m,n+1)}+\theta_{\mathrm{B}}^{(m-1,n+1)}\right]+t_1\left[\theta_{\mathrm{A}}^{(m,n)}+\theta_{\mathrm{B}}^{(m,n)}\right]=0,
	\end{equation}
	which is rewritten as
	\begin{equation}
		\label{eq5_31add} t_2\left[1+\mathrm{e}^{\mathrm{i}(\phi-k_1L)}\right]r\cdot\theta_{\mathrm{A}}^{(m,n)}+t_1\left[1+\mathrm{e}^{\mathrm{i}\phi} \right]\theta_{\mathrm{A}}^{(m,n)}=0,
	\end{equation}
	and therefore we have
	\begin{equation}
		\label{eq5_31}
		r=-\frac{t_1}{t_2}\cdot\left[ \frac{1+\mathrm{e}^{\mathrm{i}\phi}}{1+\mathrm{e}^{\mathrm{i}(\phi-k_1L)}}  \right] \quad (1+\mathrm{e}^{\mathrm{i}(\phi-k_1L)}\neq 0).
	\end{equation}
	For simplicity, in the following we let $t_1=1$, $t_2=t$. (For general cases in which $t_1\ne 1$, analogous conclusions can be reached by acknowledging the properties of the eigenvalues under scalar multiplication of a matrix, and taking $(t_1, t_2, \epsilon) \mapsto (ct_1, ct_2, c\epsilon)$.) From Eq.~\eqref{eqphiDef}, we have
	\begin{equation}
		\label{eq5_32}
		\mathrm{e}^{\mathrm{i}\phi}= \sgn(\epsilon) \frac{1+t \mathrm{e}^{\mathrm{i}k_1L}}{\lvert1+t \mathrm{e}^{\mathrm{i}k_1L}\rvert}.
	\end{equation}
	%when $\epsilon>0$, the sign of the right-hand side of Eq.~\eqref{eq5_32} is positive; when $\epsilon<0$, the sign of Eq.~\eqref{eq5_32} is negative.
	Therefore, when $\epsilon>0$,
	\begin{equation}
		\label{eq5_33}
		r=-\frac{1}{t}\left[\frac{\lvert 1+t \mathrm{e}^{\mathrm{i}k_1L}\rvert+(1+t \mathrm{e}^{\mathrm{i}k_1L})}{\lvert 1+t \mathrm{e}^{\mathrm{i}k_1L}\rvert+(t+ \mathrm{e}^{-\mathrm{i}k_1L})}\right];
	\end{equation}
	when $\epsilon<0$,
	\begin{equation}
		\label{eq5_34}
		r=-\frac{1}{t}\left[\frac{\lvert 1+t \mathrm{e}^{\mathrm{i}k_1L}\rvert-(1+t \mathrm{e}^{\mathrm{i}k_1L})}{\lvert 1+t \mathrm{e}^{\mathrm{i}k_1L}\rvert-(t+ \mathrm{e}^{-\mathrm{i}k_1L})}\right].
	\end{equation}
	Here we discuss whether the modes are localized at the boundary by enumerating the sign of $t$ and also the sign of $\epsilon$.
	
\begin{figure}[tb!] %14
		\centering
		\includegraphics[width=\linewidth]{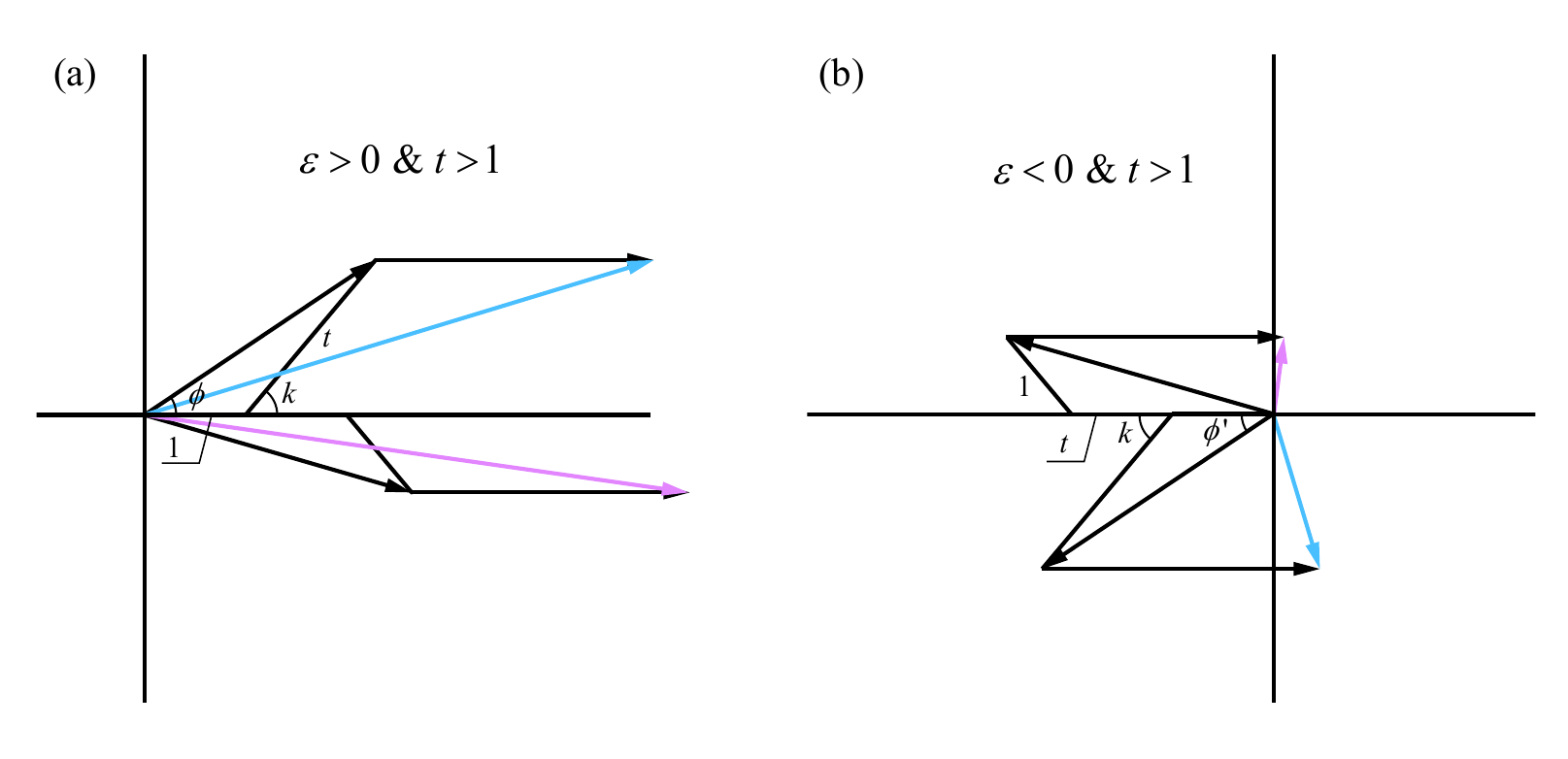}
		\caption{Geometric relation of parameters associated with the decay coefficient $r$ of the kagome frame. Here $k\equiv k_1L$. (a) When $\epsilon>0$ and $t>1$, the numerator $[\lvert 1+t \mathrm{e}^{\mathrm{i}k_1L}\rvert+(1+t \mathrm{e}^{\mathrm{i}k_1L})]$ in Eq.~\eqref{eq5_33} is indicated by the blue vector in the complex plane, and the denominator $[\lvert 1+t \mathrm{e}^{\mathrm{i}k_1L}\rvert+(t+ \mathrm{e}^{-\mathrm{i}k_1L})]$ is indicated by the purple vector. (b) When $\epsilon<0$ and $t>1$, the numerator $[\lvert 1+t \mathrm{e}^{\mathrm{i}k_1L}\rvert-(1+t \mathrm{e}^{\mathrm{i}k_1L})]$ in Eq.~\eqref{eq5_34} is indicated by the blue vector in the complex plane, and the denominator $[\lvert 1+t \mathrm{e}^{\mathrm{i}k_1L}\rvert-(t+ \mathrm{e}^{-\mathrm{i}k_1L})]$ is indicated by the purple vector. Although both illustrations implicitly assume $k = k_1L \in (0,\pi)$, for the opposite case $k = k_1L \in (-\pi,0)$, it suffices to note that $\lvert r\rvert$ is invariant under $k_1L \mapsto -k_1L$.}
		\label{Figure_14}
	\end{figure}

	\begin{itemize}
		\item Case I: $t>0$, $\epsilon>0$. Fig.~\ref{Figure_14}(a) shows the geometric relations of the parameters in the expression of $r$ (here we simply denote $k_1L$ as $k$), and by using trigonometric relations we have
		\begin{equation}
			\label{eq5_35}
			\lvert r \rvert=\frac{\sin(k-\phi)}{\sin\phi}\cdot \frac{\cos \frac{\phi}{2}}{\cos \frac{k-\phi}{2}}=\frac{\sin \frac{k-\phi}{2}}{\sin\frac{\phi}{2}}.
		\end{equation}
		When $t>1$ (which is the case shown in Fig.~\ref{Figure_14}(a)), it holds that $k-\phi<\phi$, and hence $\lvert r \rvert<1$, indicating that a mode localized at the boundary exists; on the other hand, when $0<t<1$, we have $k-\phi>\phi$, and hence $\lvert r \rvert>1$, indicating that no mode is localized at the boundary.
		\item Case II: $t>0$, $\epsilon<0$. In this case the geometric relations of the parameters in the expression of $r$ are shown in Fig.~\ref{Figure_14}(b), and we have
		\begin{equation}
			\label{eq5_36}
			\lvert r \rvert=\frac{\sin(k-\phi')}{\sin\phi'}\cdot \frac{\sin \frac{\phi'}{2}}{\sin \frac{k-\phi'}{2}}=\frac{\cos \frac{k-\phi'}{2}}{\cos\frac{\phi'}{2}}.
		\end{equation}
		When $t>1$ (as is the case in Fig.~\ref{Figure_14}(b)), it holds that $k-\phi'<\phi'$, and hence $\lvert r \rvert>1$, indicating no localization at the boundary; on the other hand, when $0<t<1$, we have $k-\phi'>\phi'$, and hence $\lvert r \rvert<1$, indicating that a mode is localized at the boundary.
		\item Case III: $t<0$, $\epsilon>0$. By a method similar to that illustrated in Fig.~\ref{Figure_14}, it is found that none of the modes exhibit localization properties.
		\item Case IV: $t<0$, $\epsilon<0$. The modes are localized at the boundaries.
	\end{itemize}
	Therefore, we arrive at the conclusion: if and only if $t_2>t_1$ and $\epsilon>0$, or $t_2<t_1$ and $\epsilon<0$, topological edge states exist at frequencies $\beta$ solved from Eq.~\eqref{eq5_24}. The condition~\eqref{eq5_25} is proved.
	
	In the above discussions, we have not yet considered the case of $r=0$ (where the modes are completely localized on the outermost layer of cells, with zero displacement in the bulk), i.e., $\mathrm{e}^{\mathrm{i}\phi}=-1$. From Eq.~\eqref{eq5_30add}, we obtain $\theta_\mathrm{B}=-\theta_\mathrm{A}$; substituting this into Eqs.~\eqref{eq5_28} and \eqref{eq5_29} yields $k_1L=0$ or $\pi$, along with the existence condition of edge states for the case of $r=0$:
	\begin{align}
		\label{eq5_36add1}
		-\epsilon &= t_1 + t_2 \quad (k_1L=0\ \text{and}\ \phi=\pi),\\
		\label{eq5_36add2}
		-\epsilon &= t_1 - t_2 \quad (k_1L=\pi\ \text{and}\ \phi=\pi).
	\end{align}
	Substituting Eq.~\eqref{eq5_36add2} into the left-hand side of the condition~\eqref{eq5_25}, it follows that $-\epsilon\cdot(t_1-t_2)=(t_1-t_2)^2>0$, and hence the solutions to Eq.~\eqref{eq5_36add2} must necessarily fall within the solution set of the condition~\eqref{eq5_25}, that is, the solutions $\beta$ to Eq.~\eqref{eq5_36add2} lie within (more precisely, lie at certain endpoints of) the yellow line segments in Fig.~\ref{Figure_11}(a), as intersections of blue curves corresponding to $\lambda_\mathrm{edge}^\mathrm{kagome}$ at $k_1L=\pi$ with the axis $\lambda=0$.
	
	However, Eq.~\eqref{eq5_36add1} does not necessarily satisfy the condition~\eqref{eq5_25} and thus requires additional discussions. From the calculations in the next subsection, we find that the solutions to Eq.~\eqref{eq5_36add1} for the edge-state frequency are also solutions for the frequency of the bulk states corresponding to $\lambda_3^\Gamma=\lambda_3^\mathrm{K}=0$. This indicates the existence of an entire bulk band (i.e., a flat band) at each of such frequencies, appearing in Fig.~\ref{Figure_11}(b) as the intersections of coincident black solid and purple dashed curves with the
	horizontal axis. When $t_1+t_2>0$, Eq.~\eqref{eq5_36add1} is produced from Eq.~\eqref{eq5_29add} by taking the negative sign, so the solutions to Eq.~\eqref{eq5_36add1} in Fig.~\ref{Figure_11}(a) are the intersections of the lower-half orange curves and the axis $\lambda=0$; when $t_1+t_2<0$, Eq.~\eqref{eq5_36add1} corresponds to Eq.~\eqref{eq5_29add} with a positive sign, and thus the solutions to Eq.~\eqref{eq5_36add1} in Fig.~\ref{Figure_11}(a) are the intersections of the upper-half orange curves and the axis $\lambda=0$. These solutions may or may not be within the range given by the condition~\eqref{eq5_25} (i.e., may or may not lie at endpoints of the yellow line segments in Fig.~\ref{Figure_11}(a)), depending on whether $\lvert t_1\rvert > \lvert t_2\rvert$ or $\lvert t_1\rvert < \lvert t_2\rvert$. We identify $\beta$ which satisfy Eq.~\eqref{eq5_36add1} but lie outside the yellow line segments with yellow stars in Fig.~\ref{Figure_11}(a). Equation~\eqref{eq5_36add1} is precisely the condition \eqref{eq5_25add} given at the beginning of this subsection. It is noted that the solutions to Eq.~\eqref{eq5_36add1} render both the numerator and denominator of Eq.~\eqref{eq5_31} zero; formally, Eq.~\eqref{eq5_31add} holds for any value of $r$. In fact, at the solutions $\beta$ to Eq.~\eqref{eq5_36add1}, edge states and bulk states coexist.
	
	\subsection{Bulk states in kagome frames}
\begin{figure}[tb!] %15
		\centering
		\includegraphics[width=\linewidth]{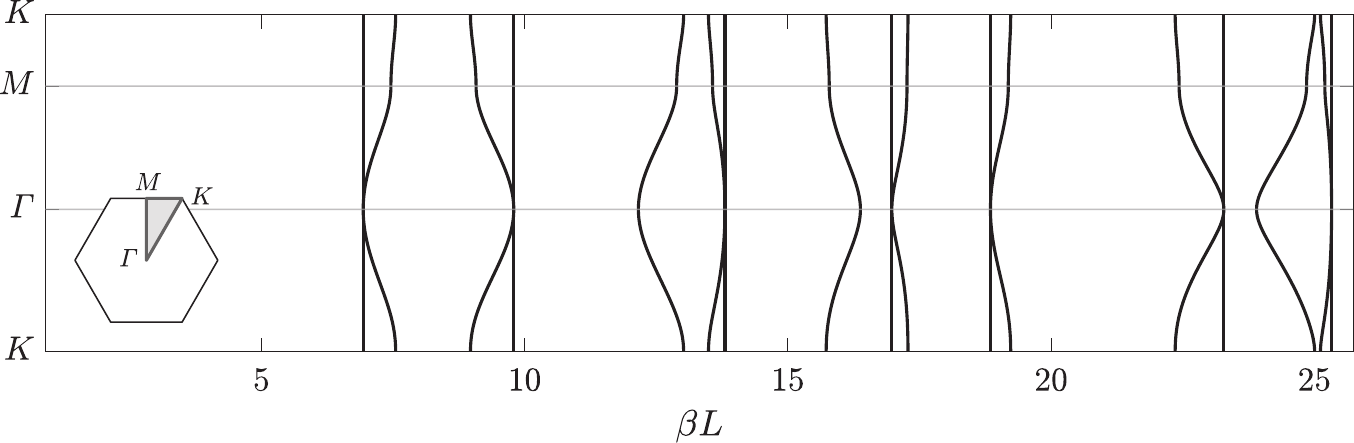}
		\caption{Band structure of the kagome frame with $(l_1, l_2) = (40\,\mathrm{mm}, 50\,\mathrm{mm})$ or $(50\,\mathrm{mm}, 40\,\mathrm{mm})$. Inset shows the first Brillouin zone.}
		\label{Figure_15}
	\end{figure}

	From the expressions in Eq.~\eqref{eq5_16_add2}, we obtain the eigenvalues of matrix $H^{\text{kagome}}_{\text{Bloch}}$ as
	\begin{align}
		\label{eq5_37}
		\lambda_{1,2}&=-\epsilon+\frac{1}{2}\left[t_1+t_2\pm\sqrt{9t_1^2+9t_2^2+t_1t_2\left(-6+8\cos{k_xL}+16\cos{\frac{k_xL}{2}}\cos{\frac{\sqrt{3}k_yL}{2}}\right)} \right],\\
		\label{eq5_38}
		\lambda_{3}&=-\epsilon-t_1-t_2.
	\end{align}
	The bulk bands are given by $\lambda_i = 0$ ($i=1$, $2$, $3$). The dispersion diagram of the bulk bands for a kagome frame with $(l_1, l_2) = (40\,\mathrm{mm}, 50\,\mathrm{mm})$ is presented in Fig.~\ref{Figure_15}.
	When $\beta$ is fixed, the upper and lower edges of the eigenvalue bands are attained at the point $\Gamma$ (i.e., $(k_x,k_y)=(0,0)$) and the point $\mathrm{K}$ (i.e., $(k_x,k_y) = \pm\bigl(\frac{2\pi}{3L},\frac{2\sqrt{3}\pi}{3L}\bigr)$) in the Brillouin zone. For point $\Gamma$,
	\begin{equation}
		\label{eq5_39}
		\lambda_{1,2}^{\Gamma} = -\epsilon + \frac{1}{2} (t_1+t_2\pm3\lvert t_1+t_2\rvert);
	\end{equation}
	for point $\mathrm{K}$,
	\begin{equation}
		\label{eq5_40}
		\lambda_{1,2}^{\mathrm{K}} = -\epsilon + \frac{1}{2} (t_1+t_2\pm3\lvert t_1-t_2\rvert);
	\end{equation}
	$\lambda_{3}^{\Gamma}=\lambda_{3}^{\mathrm{K}}$ has the same value as one of $\lambda_{1,2}^{\Gamma}$. Thus, the edges of the frequency bands for bulk states correspond to {the eigenvalues $\lambda_{1,2}^{\Gamma}$} (black solid curves in Fig.~\ref{Figure_11}(b)) {or $\lambda_{1,2}^{\mathrm{K}}$} (purple dashed curves in Fig.~\ref{Figure_11}(b)) being {equal to zero}. Comparing the values of Eqs.~\eqref{eq5_39} and \eqref{eq5_40}, it is concluded that when $t_1t_2>0$, the outermost edges of the eigenvalue bands {at any specific $\beta$} correspond to $\lambda_{1,2}^{\Gamma}$, which are black solid curves {in Fig.~\ref{Figure_11}(b)}; when $t_1t_2<0$, the outermost edges of the eigenvalue bands correspond to $\lambda_{1,2}^{\mathrm{K}}$, i.e., purple dashed curves. Here $t_1t_2=\frac{C(\beta l_1)C(\beta l_2)}{A(\beta l_1)A(\beta l_2)}$; since function $C(\beta l_i)$ is always positive for $\beta>0$, and function $A(\beta l_1)A(\beta l_2)$ changes its sign at each $\beta_0^{(n)}$, the associated wavevector (i.e., point $\Gamma$ or point $\mathrm{K}$) of the outermost bulk band edges (we exclude the flat band $\lambda_{3}$ here) will change between neighboring intervals $(\beta_0^{(n-1)},\beta_0^{(n)})$.
	
	As $\lambda_3$ is independent of the wavenumbers $k_x$ and $k_y$, the frequency bands given by $\lambda_3 = 0$ are flat bands. Remarkably, in the kagome frame, there exists a large number of such flat bands (infinitely many, in theory), due to the fact that in every frequency interval $(\beta_0^{(n-1)}, \beta_0^{(n)})$ lies one such band, as implied by the existence property of Eq.~\eqref{eq5_24} detailed earlier. These flat bands are depicted in Fig.~\ref{Figure_11}(b) as \emph{coincident} black solid and purple dashed curves intersecting the horizontal axis.
	
	\subsection{Summary of frequency ranges of states in kagome frames}
\begin{figure}[tb] %16
		\centering
		\includegraphics[width=0.8\linewidth]{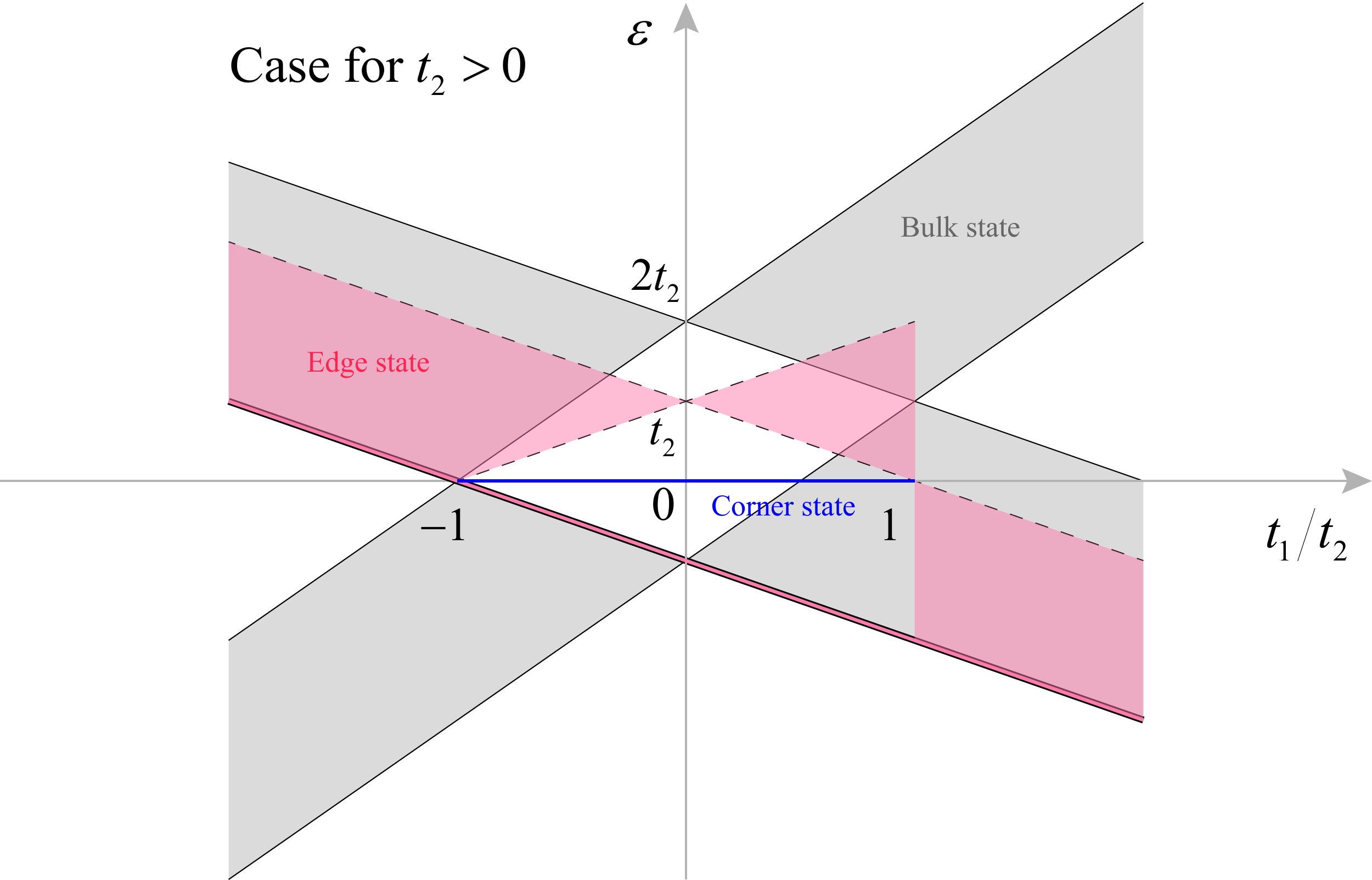}
		\caption{Spectrum of the modes in kagome lattice. Gray, pink and blue regions represent the ranges where bulk, edge and corner states exist, respectively. For the kagome frames, $\epsilon$ is defined as $-2\left[\frac{B(\beta l_1)}{A(\beta l_1)}+\frac{B(\beta l_2)}{A(\beta l_2)}\right]$, and $t_{1(2)}$ defined as $-\frac{C(\beta l_{1(2)})}{A(\beta l_{1(2)})}$. }
		\label{Figure_16}
	\end{figure}

	Finally, we summarize the frequency ranges of the corner, edge and bulk states in kagome grid-like frames, in an analogy with the tight-binding model of the breathing kagome lattice encountered in condensed matter physics \citep{xue_acoustic_2019,kagome}. {Fig.~\ref{Figure_16} illustrates the spectrum of the breathing kagome lattice (with $\epsilon$ being the energy of the eigenstates) with respect to $t_1/t_2$, where $t_1$ and $t_2$ are the intracell and intercell hopping strengths of the tight-binding lattice model; the same plot also dictates the existence of the three types of states for the kagome grid-like frames at any specified frequency $\beta$, with $\epsilon$ defined as $-2\left[\frac{B(\beta l_1)}{A(\beta l_1)}+\frac{B(\beta l_2)}{A(\beta l_2)}\right]$, and $t_{1(2)}$ defined as $-\frac{C(\beta l_{1(2)})}{A(\beta l_{1(2)})}$. }In Fig.~\ref{Figure_16}, the dark blue line denotes corner states, which exist at $\epsilon=0$ in the parameter range $-1<t_1/t_2<1$. Edge states exist in the pink regions and also at the line $\epsilon=-t_1-t_2$. Bulk states exist in the gray regions, which also contains the line $\epsilon=-t_1-t_2$.
	
	\subsection{Robustness of topological corner states in kagome frames}\label{secKagome_Robust}
		For higher-order kagome frames, although the existence criteria \eqref{eq5_24} and \eqref{eq5_25} for their edge states differ from the existence criterion Eq.~\eqref{eq5_10a} for square frames, the candidate frequency ranges of edge states, i.e., Eqs.~\eqref{eq5_10} and \eqref{eq5_24}, are identical, as well as the frequencies of corner states (Eqs.~\eqref{eq5_1} and \eqref{eq5_16}). Therefore, as long as the condition $\lvert C(\beta_tl_1) A(\beta_tl_2) \rvert = \lvert C(\beta_tl_2) A(\beta_tl_1) \lvert$ is not triggered, the corner states lie within the bandgaps of the edge states, and the existence of the corner states remains unaltered, ensuring their robustness. As demonstrated in  Fig.~\ref{Figure_16}, when the geometric parameters of the kagome grid-like frame change, the higher-order topological corner states remain within the gaps of the edge states.
	
	\section{Summary of topological states in planar frames}
	\begin{table}[t!]
		\caption{Summary of key analytical conclusions for topological modes for square and kagome frames. Here $A(\beta l)\equiv 1-\cosh \beta l \cos \beta l$, $B(\beta l)\equiv \sinh \beta l \cos \beta l-\cosh \beta l \sin \beta l$, $C(\beta l)\equiv \sinh \beta l-\sin \beta l$ are all analytical functions; $l_1$, $l_2$ are the lengths of alternately arranged beam segments and $l_0$ that of the vertical beam segments as in Figs.~\ref{Figure_1} and \ref{Figure_10}. The formulae with asterisks have been given in our previous work \citep{SUN2025105935}.}
		\centering
		\small
		\begin{tabular}{c c >{\centering}m{0.34\linewidth} >{\centering}m{0.26\linewidth}}
			\hline
			Topological frame & Type of mode & Analytical results for frequencies of modes & Existence condition for modes \tabularnewline
			\hline
			\multirow{2}{*}{Square frame} &Edge state& $2\left[\frac{B(\beta l_1)}{A(\beta l_1)}+\frac{B(\beta l_2)}{A(\beta l_2)}\right]=\pm \left\lvert\frac{C(\beta l_1)}{A(\beta l_1)}+\frac{C(\beta l_2)}{A(\beta l_2)}\exp (\mathrm{i}k_{x(y)}L)\right\rvert$& $\left\lvert\frac{C(\beta l_1)}{A(\beta l_1)} \right\rvert<\left\lvert\frac{C(\beta l_2)}{A(\beta l_2)} \right\rvert$ \tabularnewline
			\cline{2-4}
			&Corner state${}^\ast$& $2\left[\frac{B(\beta_tl_1)}{A(\beta_tl_1)}+\frac{B(\beta_tl_2)}{A(\beta_tl_2)}\right]=0$& $\left\lvert\frac{C(\beta_tl_1)}{A(\beta_tl_1)} \right\rvert<\left\lvert\frac{C(\beta_tl_2)}{A(\beta_tl_2)} \right\rvert$ \tabularnewline
			\hline
			\multirow{2}{*}{\centering Kagome frame}&Edge state& $2\left[\frac{B(\beta l_1)}{A(\beta l_1)}+\frac{B(\beta l_2)}{A(\beta l_2)}\right]=\pm \left\lvert\frac{C(\beta l_1)}{A(\beta l_1)}+\frac{C(\beta l_2)}{A(\beta l_2)}\exp (\mathrm{i}k_{\parallel }L)\right\rvert$
			& $2\left[\frac{B(\beta l_1)}{A(\beta l_1)}+\frac{B(\beta l_2)}{A(\beta l_2)}\right]\cdot \left[-\frac{C(\beta l_1)}{A(\beta l_1)}+\frac{C(\beta l_2)}{A(\beta l_2)}\right]>0$ \\ or \\ $2\left[\frac{B(\beta l_1)}{A(\beta l_1)}+\frac{B(\beta l_2)}{A(\beta l_2)}\right]=-\frac{C(\beta l_1)}{A(\beta l_1)}-\frac{C(\beta l_2)}{A(\beta l_2)}$
			\tabularnewline
			\cline{2-4}
			&Corner state${}^\ast$& $2\left[\frac{B(\beta_tl_1)}{A(\beta_tl_1)}+\frac{B(\beta_tl_2)}{A(\beta_tl_2)}\right]=0$& $\left\lvert\frac{C(\beta_tl_1)}{A(\beta_tl_1)} \right\rvert<\left\lvert\frac{C(\beta_tl_2)}{A(\beta_tl_2)} \right\rvert$\tabularnewline
%			\hline
%			Bridge-like frame&Edge state${}^\ast$& $\frac{B(\beta_tl_0)}{A(\beta_tl_0)}+\frac{B(\beta_tl_1)}{A(\beta_tl_1)}+\frac{B(\beta_tl_2)}{A(\beta_tl_2)}=0$& $\left\lvert\frac{C(\beta_tl_1)}{A(\beta_tl_1)} \right\rvert<\left\lvert\frac{C(\beta_tl_2)}{A(\beta_tl_2)} \right\rvert$ \tabularnewline
			\hline
		\end{tabular}
		\label{Table-1}
	\end{table}
	In Table~\ref{Table-1}, we briefly summarize the frequency results and existence conditions for the edge states and corner states in planar square and kagome frames, which are a class of higher-order topological continuum structures. The formulae for the corner states of the square and kagome frames have been given in our previous work \citep{SUN2025105935}. The results are concise and in analytical form, applicable to all bands from low to high frequencies, and may be used as a reference for applications.
	
	\section{Conclusions}
	In this paper, the analytical method for characterizing the higher-order topological dynamics of the continuum grid-like frames with complex spectra is proposed. We give the exact analytical expressions for all the frequencies of topological corner states, edge states, and bulk states in square and kagome frames, where spectral overlap between topological corner/edge states and bulk states occurs in such non-1D systems. Also, we present the existence conditions of edge and corner states in an analytical form, which are related to the transitions of the 2D topological phases in higher-order continuum frames. Therefore, this enables the identification of topological corner and edge states even within the densely distributed bulk states in the frequency spectrum, which is challenging with usual numerical solving techniques. Meanwhile, we find that the topological corner states must be within the bandgaps of edge states, unless topological transitions occur, demonstrating the robustness of the higher-order topological corner states from the theoretical side. In addition we verify by finite-element simulations that higher-order topological corner states persist robustly under perturbation of various types of defects. Furthermore, we extend our grid-like frame to demonstrate the corner states at the interface of topological heterostructures. The clear identification of mode distributions in the frequency spectra of grid-like frame structures contributes to the rigorous theoretical analysis of higher-order topological dynamics in continuum systems, and the demonstration on the robustness of the topological states facilitates potential applications of frame structures such as robust waveguides and safety assessment in engineering \citep{VASILIEV2006182,VASILIEV20121117,KIAKOJOURI2020110061,zhang_dirac_2020,jiangetalpnas2023,wang_topological_2026}.
	
	\section*{CRediT authorship contribution statement}
	\textbf{Yimeng Sun: }Conceptualization, Methodology, Software, Formal analysis, Validation, Writing -- original draft, Writing -- review \& editing, Visualization. \textbf{Jiacheng Xing: }Conceptualization, Methodology, Software, Formal analysis, Validation, Writing -- original draft, Writing -- review \& editing, Visualization. \textbf{Li-Hua Shao: }Conceptualization, Resources, Validation, Writing -- review \& editing. \textbf{Jianxiang Wang: }Conceptualization, Methodology, Formal analysis, Validation, Resources, Supervision, Writing -- review \& editing, Funding acquisition.
	\section*{Declaration of Competing Interest}
	The authors declare that they have no known competing financial interests or personal relationships that could have appeared to influence the work reported in this paper.
	\section*{Data availability}
	Data will be made available on request.
	\section*{Acknowledgements}
	Y. S., J. X., and J. W. thank the National Natural Science Foundation of China (Grant No.\ 12232001) for support of this work. We thank Dr.\ Hao Qiu from Suzhou Laboratory for helpful discussions.
	
	% \appendix
	% \renewcommand{\thefigure}{\arabic{figure}}

	%\clearpage
	
	
	
	\bibliographystyle{elsarticle-num-names}
	\bibliography{sunyimeng}

	
	%% \begin{thebibliography}{00}
		%% \bibitem[ ()]{}
		%%
		%% \end{thebibliography}
\end{document}